\magnification=\magstep1 
\font\bigbfont=cmbx10 scaled\magstep1
\font\bigifont=cmti10 scaled\magstep1
\font\bigrfont=cmr10 scaled\magstep1
\vsize = 23.5 truecm
\hsize = 15.5 truecm
\hoffset = .2truein
\baselineskip = 14 truept
\overfullrule = 0pt
\parskip = 3 truept
\def\bi{\text {\bf i}}
\def\bj{\text {\bf j}}
\def\bk{\text {\bf k}}

\def\fmtname{AmS-TeX}

\def\fmtversion{2.1}
\catcode`\@=11
\ifx\amstexloaded@\relax\catcode`\@=\active
  \endinput\else\let\amstexloaded@\relax\fi
\newlinechar=`\^^J
\def\W@{\immediate\write\sixt@@n}
\def\CR@{\W@{^^J\fmtname - Version \fmtversion^^J%
COPYRIGHT 1985, 1990, 1991 - AMERICAN MATHEMATICAL SOCIETY^^J%
Use of this macro package is not restricted provided^^J%
each use is acknowledged upon publication.^^J}}
\CR@ \everyjob{\CR@}
\message{Loading definitions for}
\message{misc utility macros,}
\toksdef\toks@@=2
\long\def\rightappend@#1\to#2{\toks@{\\{#1}}\toks@@
 =\expandafter{#2}\xdef#2{\the\toks@@\the\toks@}\toks@{}\toks@@{}}
\def\alloclist@{}
\newif\ifalloc@
\def\showallocations{{\def\\{\immediate\write\m@ne}\alloclist@}\alloc@true}
\def\alloc@#1#2#3#4#5{\global\advance\count1#1by\@ne
 \ch@ck#1#4#2\allocationnumber=\count1#1
 \global#3#5=\allocationnumber
 \edef\next@{\string#5=\string#2\the\allocationnumber}%
 \expandafter\rightappend@\next@\to\alloclist@}
\newcount\count@@
\newcount\count@@@
\def\FN@{\futurelet\next}
\def\DN@{\def\next@}
\def\DNii@{\def\nextii@}
\def\RIfM@{\relax\ifmmode}
\def\RIfMIfI@{\relax\ifmmode\ifinner}
\def\setboxz@h{\setbox\z@\hbox}
\def\wdz@{\wd\z@}
\def\boxz@{\box\z@}
\def\setbox@ne{\setbox\@ne}
\def\wd@ne{\wd\@ne}
\def\iterate{\body\expandafter\iterate\else\fi}
\def\err@#1{\errmessage{AmS-TeX error: #1}}
\newhelp\defaulthelp@{Sorry, I already gave what help I could...^^J
Maybe you should try asking a human?^^J
An error might have occurred before I noticed any problems.^^J
``If all else fails, read the instructions.''}
\def\Err@{\errhelp\defaulthelp@\err@}
\def\eat@#1{}
\def\in@#1#2{\def\in@@##1#1##2##3\in@@{\ifx\in@##2\in@false\else\in@true\fi}%
 \in@@#2#1\in@\in@@}
\newif\ifin@
\def\space@.{\futurelet\space@\relax}
\space@. %
\newhelp\athelp@
{Only certain combinations beginning with @ make sense to me.^^J
Perhaps you wanted \string\@\space for a printed @?^^J
I've ignored the character or group after @.}
{\catcode`\~=\active % just in case
 \lccode`\~=`\@ \lowercase{\gdef~{\FN@\at@}}}
\def\at@{\let\next@\at@@
 \ifcat\noexpand\next a\else\ifcat\noexpand\next0\else
 \ifcat\noexpand\next\relax\else
   \let\next\at@@@\fi\fi\fi
 \next@}
\def\at@@#1{\expandafter
 \ifx\csname\space @\string#1\endcsname\relax
  \expandafter\at@@@ \else
  \csname\space @\string#1\expandafter\endcsname\fi}
\def\at@@@#1{\errhelp\athelp@ \err@{\Invalid@@ @}}%%
\def\atdef@#1{\expandafter\def\csname\space @\string#1\endcsname}%%
\newhelp\defahelp@{If you typed \string\define\space cs instead of
\string\define\string\cs\space^^J
I've substituted an inaccessible control sequence so that your^^J
definition will be completed without mixing me up too badly.^^J
If you typed \string\define{\string\cs} the inaccessible control sequence^^J
was defined to be \string\cs, and the rest of your^^J
definition appears as input.}
\newhelp\defbhelp@{I've ignored your definition, because it might^^J
conflict with other uses that are important to me.}
\def\define{\FN@\define@}
\def\define@{\ifcat\noexpand\next\relax
 \expandafter\define@@\else\errhelp\defahelp@                               %1
 \err@{\string\define\space must be followed by a control
 sequence}\expandafter\def\expandafter\nextii@\fi}                          %2
\def\undefined@@@@@@@@@@{}
\def\preloaded@@@@@@@@@@{}
\def\next@@@@@@@@@@{}
\def\define@@#1{\ifx#1\relax\errhelp\defbhelp@                              %1
 \err@{\string#1\space is already defined}\DN@{\DNii@}\else
 \expandafter\ifx\csname\expandafter\eat@\string                            %2
 #1@@@@@@@@@@\endcsname\undefined@@@@@@@@@@\errhelp\defbhelp@
 \err@{\string#1\space can't be defined}\DN@{\DNii@}\else
 \expandafter\ifx\csname\expandafter\eat@\string#1\endcsname\relax          %3
 \global\let#1\undefined\DN@{\def#1}\else\errhelp\defbhelp@
 \err@{\string#1\space is already defined}\DN@{\DNii@}\fi
 \fi\fi\next@}

\def\predefine#1#2{\let#1#2}
\def\undefine#1{\let#1\undefined}
\message{page layout,}
\newdimen\captionwidth@
\captionwidth@\hsize
\advance\captionwidth@-1.5in
\def\pagewidth#1{\hsize#1\relax
 \captionwidth@\hsize\advance\captionwidth@-1.5in}

\def\hcorrection#1{\advance\hoffset#1\relax}
\def\vcorrection#1{\advance\voffset#1\relax}
\message{accents/punctuation,}

\let\graveaccent\`
\let\acuteaccent\'
\let\tildeaccent\~
\let\hataccent\^
\let\underscore\_
\let\B\=
\let\D\.
\let\ic@\/
\def\/{\unskip\ic@}
\def\textfonti{\the\textfont\@ne}
\def\t#1#2{{\edef\next@{\the\font}\textfonti\accent"7F \next@#1#2}}
\def~{\unskip\nobreak\ \ignorespaces}
\def\.{.\spacefactor\@m}
\atdef@;{\leavevmode\null;}
\atdef@:{\leavevmode\null:}
\atdef@?{\leavevmode\null?}
\edef\@{\string @}
\def\relaxnext@{\let\next\relax}
\atdef@-{\relaxnext@\leavevmode
 \DN@{\ifx\next-\DN@-{\FN@\nextii@}\else
  \DN@{\leavevmode\hbox{-}}\fi\next@}%
 \DNii@{\ifx\next-\DN@-{\leavevmode\hbox{---}}\else
  \DN@{\leavevmode\hbox{--}}\fi\next@}%
 \FN@\next@}
\def\srdr@{\kern.16667em}
\def\drsr@{\kern.02778em}
\def\sldl@{\drsr@}
\def\dlsl@{\srdr@}
\atdef@"{\unskip\relaxnext@
 \DN@{\ifx\next\space@\DN@. {\FN@\nextii@}\else
  \DN@.{\FN@\nextii@}\fi\next@.}%
 \DNii@{\ifx\next`\DN@`{\FN@\nextiii@}\else
  \ifx\next\lq\DN@\lq{\FN@\nextiii@}\else
  \DN@####1{\FN@\nextiv@}\fi\fi\next@}%
 \def\nextiii@{\ifx\next`\DN@`{\sldl@``}\else\ifx\next\lq
  \DN@\lq{\sldl@``}\else\DN@{\dlsl@`}\fi\fi\next@}%
 \def\nextiv@{\ifx\next'\DN@'{\srdr@''}\else
  \ifx\next\rq\DN@\rq{\srdr@''}\else\DN@{\drsr@'}\fi\fi\next@}%
 \FN@\next@}

\def\textfontii{\the\textfont\tw@}
\def\lbrace@{\delimiter"4266308 }
\def\rbrace@{\delimiter"5267309 }
\def\{{\RIfM@\lbrace@\else{\textfontii f}\spacefactor\@m\fi}
\def\}{\RIfM@\rbrace@\else
 \let\@sf\empty\ifhmode\edef\@sf{\spacefactor\the\spacefactor}\fi
 {\textfontii g}\@sf\relax\fi}
\let\lbrace\{
\let\rbrace\}
\def\AmSTeX{{\textfontii A\kern-.1667em%
  \lower.5ex\hbox{M}\kern-.125emS}-\TeX}
\message{line and page breaks,}
\def\vmodeerr@#1{\Err@{\string#1\space not allowed between paragraphs}}
\def\mathmodeerr@#1{\Err@{\string#1\space not allowed in math mode}}
\def\linebreak{\RIfM@\mathmodeerr@\linebreak\else
 \ifhmode\unskip\unkern\break\else\vmodeerr@\linebreak\fi\fi}

\newskip\saveskip@
\def\allowlinebreak{\RIfM@\mathmodeerr@\allowlinebreak\else
 \ifhmode\saveskip@\lastskip\unskip
 \allowbreak\ifdim\saveskip@>\z@\hskip\saveskip@\fi
 \else\vmodeerr@\allowlinebreak\fi\fi}
\def\nolinebreak{\RIfM@\mathmodeerr@\nolinebreak\else
 \ifhmode\saveskip@\lastskip\unskip
 \nobreak\ifdim\saveskip@>\z@\hskip\saveskip@\fi
 \else\vmodeerr@\nolinebreak\fi\fi}
\def\newline{\relaxnext@
 \DN@{\RIfM@\expandafter\mathmodeerr@\expandafter\newline\else
  \ifhmode\ifx\next\par\else
  \expandafter\unskip\expandafter\null\expandafter\hfill\expandafter\break\fi
  \else
  \expandafter\vmodeerr@\expandafter\newline\fi\fi}%
 \FN@\next@}
\def\dmatherr@#1{\Err@{\string#1\space not allowed in display math mode}}
\def\nondmatherr@#1{\Err@{\string#1\space not allowed in non-display math
 mode}}
\def\onlydmatherr@#1{\Err@{\string#1\space allowed only in display math mode}}
\def\nonmatherr@#1{\Err@{\string#1\space allowed only in math mode}}
\def\mathbreak{\RIfMIfI@\break\else
 \dmatherr@\mathbreak\fi\else\nonmatherr@\mathbreak\fi}
\def\nomathbreak{\RIfMIfI@\nobreak\else
 \dmatherr@\nomathbreak\fi\else\nonmatherr@\nomathbreak\fi}
\def\allowmathbreak{\RIfMIfI@\allowbreak\else
 \dmatherr@\allowmathbreak\fi\else\nonmatherr@\allowmathbreak\fi}
\def\pagebreak{\RIfM@
 \ifinner\nondmatherr@\pagebreak\else\postdisplaypenalty-\@M\fi
 \else\ifvmode\removelastskip\break\else\vadjust{\break}\fi\fi}
\def\nopagebreak{\RIfM@
 \ifinner\nondmatherr@\nopagebreak\else\postdisplaypenalty\@M\fi
 \else\ifvmode\nobreak\else\vadjust{\nobreak}\fi\fi}
\def\nonvmodeerr@#1{\Err@{\string#1\space not allowed within a paragraph
 or in math}}
\def\vnonvmode@#1#2{\relaxnext@\DNii@{\ifx\next\par\DN@{#1}\else
 \DN@{#2}\fi\next@}%
 \ifvmode\DN@{#1}\else
 \DN@{\FN@\nextii@}\fi\next@}
\def\newpage{\vnonvmode@{\vfill\break}{\nonvmodeerr@\newpage}}
\def\smallpagebreak{\vnonvmode@\smallbreak{\nonvmodeerr@\smallpagebreak}}
\def\medpagebreak{\vnonvmode@\medbreak{\nonvmodeerr@\medpagebreak}}
\def\bigpagebreak{\vnonvmode@\bigbreak{\nonvmodeerr@\bigpagebreak}}
\def\NoBlackBoxes{\global\overfullrule\z@}
\def\BlackBoxes{\global\overfullrule5\p@}
\def\Invalid@#1{\def#1{\Err@{\Invalid@@\string#1}}}
\def\Invalid@@{Invalid use of }
\message{figures,}
\Invalid@\caption
\Invalid@\captionwidth
\newdimen\smallcaptionwidth@
\def\topspace{\mid@false\ins@}
\def\midspace{\mid@true\ins@}
\newif\ifmid@
\def\captionfont@{}
\def\ins@#1{\relaxnext@\allowbreak
 \smallcaptionwidth@\captionwidth@\gdef\thespace@{#1}%
 \DN@{\ifx\next\space@\DN@. {\FN@\nextii@}\else
  \DN@.{\FN@\nextii@}\fi\next@.}%
 \DNii@{\ifx\next\caption\DN@\caption{\FN@\nextiii@}%
  \else\let\next@\nextiv@\fi\next@}%
 \def\nextiv@{\vnonvmode@
  {\ifmid@\expandafter\midinsert\else\expandafter\topinsert\fi
   \vbox to\thespace@{}\endinsert}
  {\ifmid@\nonvmodeerr@\midspace\else\nonvmodeerr@\topspace\fi}}%
 \def\nextiii@{\ifx\next\captionwidth\expandafter\nextv@
  \else\expandafter\nextvi@\fi}%
 \def\nextv@\captionwidth##1##2{\smallcaptionwidth@##1\relax\nextvi@{##2}}%
 \def\nextvi@##1{\def\thecaption@{\captionfont@##1}%
  \DN@{\ifx\next\space@\DN@. {\FN@\nextvii@}\else
   \DN@.{\FN@\nextvii@}\fi\next@.}%
  \FN@\next@}%
 \def\nextvii@{\vnonvmode@
  {\ifmid@\expandafter\midinsert\else
  \expandafter\topinsert\fi\vbox to\thespace@{}\nobreak\smallskip
  \setboxz@h{\noindent\ignorespaces\thecaption@\unskip}%
  \ifdim\wdz@>\smallcaptionwidth@\centerline{\vbox{\hsize\smallcaptionwidth@
   \noindent\ignorespaces\thecaption@\unskip}}%
  \else\centerline{\boxz@}\fi\endinsert}
  {\ifmid@\nonvmodeerr@\midspace
  \else\nonvmodeerr@\topspace\fi}}%
 \FN@\next@}
\message{comments,}
\def\newcodes@{\catcode`\\12\catcode`\{12\catcode`\}12\catcode`\#12%
 \catcode`\%12\relax}
\def\oldcodes@{\catcode`\\0\catcode`\{1\catcode`\}2\catcode`\#6%
 \catcode`\%14\relax}
\def\comment{\newcodes@\endlinechar=10 \comment@}
{\lccode`\0=`\\
\lowercase{\gdef\comment@#1^^J{\comment@@#10endcomment\comment@@@}%
\gdef\comment@@#10endcomment{\FN@\comment@@@}%
\gdef\comment@@@#1\comment@@@{\ifx\next\comment@@@\let\next\comment@
 \else\def\next{\oldcodes@\endlinechar=`\^^M\relax}%
 \fi\next}}}
\def\pr@m@s{\ifx'\next\DN@##1{\prim@s}\else\let\next@\egroup\fi\next@}
\def\prime{{\null\prime@\null}}
\mathchardef\prime@="0230
\let\dsize\displaystyle

\let\ssize\scriptstyle

\message{math spacing,}
\def\,{\RIfM@\mskip\thinmuskip\relax\else\kern.16667em\fi}
\def\!{\RIfM@\mskip-\thinmuskip\relax\else\kern-.16667em\fi}
\let\thinspace\,
\let\negthinspace\!
\def\medspace{\RIfM@\mskip\medmuskip\relax\else\kern.222222em\fi}
\def\negmedspace{\RIfM@\mskip-\medmuskip\relax\else\kern-.222222em\fi}
\def\thickspace{\RIfM@\mskip\thickmuskip\relax\else\kern.27777em\fi}
\let\;\thickspace
\def\negthickspace{\RIfM@\mskip-\thickmuskip\relax\else
 \kern-.27777em\fi}
\atdef@,{\RIfM@\mskip.1\thinmuskip\else\leavevmode\null,\fi}
\atdef@!{\RIfM@\mskip-.1\thinmuskip\else\leavevmode\null!\fi}
\atdef@.{\RIfM@&&\else\leavevmode.\spacefactor3000 \fi}
\def\and{\DOTSB\;\mathchar"3026 \;}

\message{fractions,}
\def\frac#1#2{{#1\over#2}}

\newdimen\ex@
\ex@.2326ex
\Invalid@\thickness
\def\thickfrac{\relaxnext@
 \DN@{\ifx\next\thickness\let\next@\nextii@\else
 \DN@{\nextii@\thickness1}\fi\next@}%
 \DNii@\thickness##1##2##3{{##2\above##1\ex@##3}}%
 \FN@\next@}

\def\thickfracwithdelims#1#2{\relaxnext@\def\ldelim@{#1}\def\rdelim@{#2}%
 \DN@{\ifx\next\thickness\let\next@\nextii@\else
 \DN@{\nextii@\thickness1}\fi\next@}%
 \DNii@\thickness##1##2##3{{##2\abovewithdelims
 \ldelim@\rdelim@##1\ex@##3}}%
 \FN@\next@}

\def\:{\nobreak\hskip.1111em\mathpunct{}\nonscript\mkern-\thinmuskip{:}\hskip
 .3333emplus.0555em\relax}
\def\snug{\unskip\kern-\mathsurround}
\message{smash commands,}
\def\topsmash{\top@true\bot@false\smash@}
\def\botsmash{\top@false\bot@true\smash@}
\newif\iftop@
\newif\ifbot@
\def\smash{\top@true\bot@true\smash@}
\def\smash@{\RIfM@\expandafter\mathpalette\expandafter\mathsm@sh\else
 \expandafter\makesm@sh\fi}
\def\finsm@sh{\iftop@\ht\z@\z@\fi\ifbot@\dp\z@\z@\fi\leavevmode\boxz@}
\message{large operator symbols,}
\def\LimitsOnSums{\global\let\slimits@\displaylimits}
\def\NoLimitsOnSums{\global\let\slimits@\nolimits}
\LimitsOnSums
\mathchardef\coprod@="1360       \def\coprod{\DOTSB\coprod@\slimits@}
\mathchardef\bigvee@="1357       \def\bigvee{\DOTSB\bigvee@\slimits@}
\mathchardef\bigwedge@="1356     \def\bigwedge{\DOTSB\bigwedge@\slimits@}
\mathchardef\biguplus@="1355     \def\biguplus{\DOTSB\biguplus@\slimits@}
\mathchardef\bigcap@="1354       \def\bigcap{\DOTSB\bigcap@\slimits@}
\mathchardef\bigcup@="1353       \def\bigcup{\DOTSB\bigcup@\slimits@}
\mathchardef\prod@="1351         \def\prod{\DOTSB\prod@\slimits@}
\mathchardef\sum@="1350          \def\sum{\DOTSB\sum@\slimits@}
\mathchardef\bigotimes@="134E    \def\bigotimes{\DOTSB\bigotimes@\slimits@}
\mathchardef\bigoplus@="134C     \def\bigoplus{\DOTSB\bigoplus@\slimits@}
\mathchardef\bigodot@="134A      \def\bigodot{\DOTSB\bigodot@\slimits@}
\mathchardef\bigsqcup@="1346     \def\bigsqcup{\DOTSB\bigsqcup@\slimits@}
\message{integrals,}
\def\LimitsOnInts{\global\let\ilimits@\displaylimits}
\def\NoLimitsOnInts{\global\let\ilimits@\nolimits}
\NoLimitsOnInts
\def\int{\DOTSI\intop\ilimits@}
\def\oint{\DOTSI\ointop\ilimits@}
\def\intic@{\mathchoice{\hskip.5em}{\hskip.4em}{\hskip.4em}{\hskip.4em}}
\def\negintic@{\mathchoice
 {\hskip-.5em}{\hskip-.4em}{\hskip-.4em}{\hskip-.4em}}
\def\intkern@{\mathchoice{\!\!\!}{\!\!}{\!\!}{\!\!}}
\def\intdots@{\mathchoice{\plaincdots@}
 {{\cdotp}\mkern1.5mu{\cdotp}\mkern1.5mu{\cdotp}}
 {{\cdotp}\mkern1mu{\cdotp}\mkern1mu{\cdotp}}
 {{\cdotp}\mkern1mu{\cdotp}\mkern1mu{\cdotp}}}
\newcount\intno@
\def\iint{\DOTSI\intno@\tw@\FN@\ints@}
\def\iiint{\DOTSI\intno@\thr@@\FN@\ints@}
\def\iiiint{\DOTSI\intno@4 \FN@\ints@}
\def\idotsint{\DOTSI\intno@\z@\FN@\ints@}
\def\ints@{\findlimits@\ints@@}
\newif\iflimtoken@
\newif\iflimits@
\def\findlimits@{\limtoken@true\ifx\next\limits\limits@true
 \else\ifx\next\nolimits\limits@false\else
 \limtoken@false\ifx\ilimits@\nolimits\limits@false\else
 \ifinner\limits@false\else\limits@true\fi\fi\fi\fi}
\def\multint@{\int\ifnum\intno@=\z@\intdots@                                %1
 \else\intkern@\fi                                                          %2
 \ifnum\intno@>\tw@\int\intkern@\fi                                         %3
 \ifnum\intno@>\thr@@\int\intkern@\fi                                       %4
 \int}                                                                      %5
\def\multintlimits@{\intop\ifnum\intno@=\z@\intdots@\else\intkern@\fi
 \ifnum\intno@>\tw@\intop\intkern@\fi
 \ifnum\intno@>\thr@@\intop\intkern@\fi\intop}
\def\ints@@{\iflimtoken@                                                    %1
 \def\ints@@@{\iflimits@\negintic@\mathop{\intic@\multintlimits@}\limits    %2
  \else\multint@\nolimits\fi                                                %3
  \eat@}                                                                    %4
 \else                                                                      %5
 \def\ints@@@{\iflimits@\negintic@
  \mathop{\intic@\multintlimits@}\limits\else
  \multint@\nolimits\fi}\fi\ints@@@}
\def\LimitsOnNames{\global\let\nlimits@\displaylimits}
\def\NoLimitsOnNames{\global\let\nlimits@\nolimits@}
\LimitsOnNames
\def\nolimits@{\relaxnext@
 \DN@{\ifx\next\limits\DN@\limits{\nolimits}\else
  \let\next@\nolimits\fi\next@}%
 \FN@\next@}
\message{operator names,}
\def\newmcodes@{\mathcode`\'"27\mathcode`\*"2A\mathcode`\."613A%
 \mathcode`\-"2D\mathcode`\/"2F\mathcode`\:"603A }
\def\operatorname#1{\mathop{\newmcodes@\kern\z@\fam\z@#1}\nolimits@}
\def\operatornamewithlimits#1{\mathop{\newmcodes@\kern\z@\fam\z@#1}\nlimits@}
\def\qopname@#1{\mathop{\fam\z@#1}\nolimits@}
\def\qopnamewl@#1{\mathop{\fam\z@#1}\nlimits@}
\def\arccos{\qopname@{arccos}}
\def\arcsin{\qopname@{arcsin}}
\def\arctan{\qopname@{arctan}}
\def\arg{\qopname@{arg}}
\def\cos{\qopname@{cos}}
\def\cosh{\qopname@{cosh}}
\def\cot{\qopname@{cot}}
\def\coth{\qopname@{coth}}
\def\csc{\qopname@{csc}}
\def\deg{\qopname@{deg}}
\def\det{\qopnamewl@{det}}
\def\dim{\qopname@{dim}}
\def\exp{\qopname@{exp}}
\def\gcd{\qopnamewl@{gcd}}
\def\hom{\qopname@{hom}}
\def\inf{\qopnamewl@{inf}}
\def\injlim{\qopnamewl@{inj\,lim}}
\def\ker{\qopname@{ker}}
\def\lg{\qopname@{lg}}
\def\lim{\qopnamewl@{lim}}
\def\liminf{\qopnamewl@{lim\,inf}}
\def\limsup{\qopnamewl@{lim\,sup}}
\def\ln{\qopname@{ln}}
\def\log{\qopname@{log}}
\def\max{\qopnamewl@{max}}
\def\min{\qopnamewl@{min}}
\def\Pr{\qopnamewl@{Pr}}
\def\projlim{\qopnamewl@{proj\,lim}}
\def\sec{\qopname@{sec}}
\def\sin{\qopname@{sin}}
\def\sinh{\qopname@{sinh}}
\def\sup{\qopnamewl@{sup}}
\def\tan{\qopname@{tan}}
\def\tanh{\qopname@{tanh}}
\def\varinjlim{\mathop{\vtop{\ialign{##\crcr
 \hfil\rm lim\hfil\crcr\noalign{\nointerlineskip}\rightarrowfill\crcr
 \noalign{\nointerlineskip\kern-\ex@}\crcr}}}}
\def\varprojlim{\mathop{\vtop{\ialign{##\crcr
 \hfil\rm lim\hfil\crcr\noalign{\nointerlineskip}\leftarrowfill\crcr
 \noalign{\nointerlineskip\kern-\ex@}\crcr}}}}
\def\varliminf{\mathop{\underline{\vrule height\z@ depth.2exwidth\z@
 \hbox{\rm lim}}}}

\newdimen\buffer@
\buffer@\fontdimen13 \tenex
\newdimen\buffer
\buffer\buffer@

\def\ResetBuffer{\fontdimen13 \tenex\buffer@\global\buffer\buffer@}
\def\shave#1{\mathop{\hbox{$\m@th\fontdimen13 \tenex\z@                     %1
 \displaystyle{#1}$}}\fontdimen13 \tenex\buffer}

\message{multilevel sub/superscripts,}
\Invalid@\\
\def\Let@{\relax\iffalse{\fi\let\\=\cr\iffalse}\fi}
\Invalid@\vspace
\def\vspace@{\def\vspace##1{\crcr\noalign{\vskip##1\relax}}}
\def\multilimits@{\bgroup\vspace@\Let@
 \baselineskip\fontdimen10 \scriptfont\tw@
 \advance\baselineskip\fontdimen12 \scriptfont\tw@
 \lineskip\thr@@\fontdimen8 \scriptfont\thr@@
 \lineskiplimit\lineskip
 \vbox\bgroup\ialign\bgroup\hfil$\m@th\scriptstyle{##}$\hfil\crcr}
\def\Sb{_\multilimits@}
\def\endSb{\crcr\egroup\egroup\egroup}
\def\Sp{^\multilimits@}

\def\spreadlines#1{\RIfMIfI@\onlydmatherr@\spreadlines\else
 \openup#1\relax\fi\else\onlydmatherr@\spreadlines\fi}
\def\Mathstrut@{\copy\Mathstrutbox@}
\newbox\Mathstrutbox@
\setbox\Mathstrutbox@\null
\setboxz@h{$\m@th($}
\ht\Mathstrutbox@\ht\z@
\dp\Mathstrutbox@\dp\z@
\message{matrices,}
\newdimen\spreadmlines@
\def\spreadmatrixlines#1{\RIfMIfI@
 \onlydmatherr@\spreadmatrixlines\else
 \spreadmlines@#1\relax\fi\else\onlydmatherr@\spreadmatrixlines\fi}
\def\format{\crcr\egroup\iffalse{\fi\ifnum`}=0 \fi\format@}
\newtoks\hashtoks@
\hashtoks@{#}
\def\format@#1\\{\def\preamble@{#1}%
 \def\l{$\m@th\the\hashtoks@$\hfil}%
 \def\c{\hfil$\m@th\the\hashtoks@$\hfil}%
 \def\r{\hfil$\m@th\the\hashtoks@$}%
 \edef\preamble@@{\preamble@}\ifnum`{=0 \fi\iffalse}\fi
 \ialign\bgroup\span\preamble@@\crcr}
\def\smallmatrix{\null\,\vcenter\bgroup\vspace@\Let@
 \baselineskip9\ex@\lineskip\ex@
 \ialign\bgroup\hfil$\m@th\scriptstyle{##}$\hfil&&\thickspace\hfil
 $\m@th\scriptstyle{##}$\hfil\crcr}
\def\endsmallmatrix{\crcr\egroup\egroup\,}

\newmuskip\dotsspace@
\dotsspace@1.5mu
\def\strip@#1 {#1}
\def\spacehdots#1\for#2{\multispan{#2}\xleaders
 \hbox{$\m@th\mkern\strip@#1 \dotsspace@.\mkern\strip@#1 \dotsspace@$}\hfill}
\def\hdotsfor#1{\spacehdots\@ne\for{#1}}
\def\multispan@#1{\omit\mscount#1\unskip\loop\ifnum\mscount>\@ne\sp@n\repeat}
\def\spaceinnerhdots#1\for#2\after#3{\multispan@{\strip@#2 }#3\xleaders
 \hbox{$\m@th\mkern\strip@#1 \dotsspace@.\mkern\strip@#1 \dotsspace@$}\hfill}
\def\innerhdotsfor#1\after#2{\spaceinnerhdots\@ne\for#1\after{#2}}
\def\endcases{\endmatrix\right.\egroup}
\message{multiline displays,}
\newif\ifinany@
\newif\ifinalign@
\newif\ifingather@
\def\strut@{\copy\strutbox@}
\newbox\strutbox@
\setbox\strutbox@\hbox{\vrule height8\p@ depth3\p@ width\z@}
\def\topaligned{\null\,\vtop\aligned@}
\def\botaligned{\null\,\vbox\aligned@}
\def\aligned{\null\,\vcenter\aligned@}
\def\aligned@{\bgroup\vspace@\Let@
 \ifinany@\else\openup\jot\fi\ialign
 \bgroup\hfil\strut@$\m@th\displaystyle{##}$&
 $\m@th\displaystyle{{}##}$\hfil\crcr}
\def\endaligned{\crcr\egroup\egroup}

\def\alignedat#1{\null\,\vcenter\bgroup\doat@{#1}\vspace@\Let@
 \ifinany@\else\openup\jot\fi\ialign\bgroup\span\preamble@@\crcr}
\newcount\atcount@
\def\doat@#1{\toks@{\hfil\strut@$\m@th
 \displaystyle{\the\hashtoks@}$&$\m@th\displaystyle
 {{}\the\hashtoks@}$\hfil}%                                                 %1
 \atcount@#1\relax\advance\atcount@\m@ne                                    %2
 \loop\ifnum\atcount@>\z@\toks@=\expandafter{\the\toks@&\hfil$\m@th
 \displaystyle{\the\hashtoks@}$&$\m@th
 \displaystyle{{}\the\hashtoks@}$\hfil}\advance
  \atcount@\m@ne\repeat                                                     %3
 \xdef\preamble@{\the\toks@}\xdef\preamble@@{\preamble@}}

\def\gathered{\null\,\vcenter\bgroup\vspace@\Let@
 \ifinany@\else\openup\jot\fi\ialign
 \bgroup\hfil\strut@$\m@th\displaystyle{##}$\hfil\crcr}
\def\endgathered{\crcr\egroup\egroup}
\newif\iftagsleft@
\def\TagsOnLeft{\global\tagsleft@true}
\def\TagsOnRight{\global\tagsleft@false}
\TagsOnLeft
\newif\ifmathtags@
\def\TagsAsMath{\global\mathtags@true}
\def\TagsAsText{\global\mathtags@false}
\TagsAsText
\def\tagform@#1{\hbox{\rm(\ignorespaces#1\unskip)}}
\def\thetag{\leavevmode\tagform@}
\def\tag#1$${\iftagsleft@\leqno\else\eqno\fi                                %1
 \maketag@#1\maketag@                                                       %2
 $$}                                                                        %3
\def\maketag@{\FN@\maketag@@}
\def\maketag@@{\ifx\next"\expandafter\maketag@@@\else\expandafter\maketag@@@@
 \fi}
\def\maketag@@@"#1"#2\maketag@{\hbox{\rm#1}}                                %1
\def\maketag@@@@#1\maketag@{\ifmathtags@\tagform@{$\m@th#1$}\else
 \tagform@{#1}\fi}
\interdisplaylinepenalty\@M
\def\allowdisplaybreaks{\RIfMIfI@
 \onlydmatherr@\allowdisplaybreaks\else
 \interdisplaylinepenalty\z@\fi\else\onlydmatherr@\allowdisplaybreaks\fi}
\Invalid@\allowdisplaybreak
\Invalid@\displaybreak
\Invalid@\intertext
\def\allowdisplaybreak@{\def\allowdisplaybreak{\crcr\noalign{\allowbreak}}}
\def\displaybreak@{\def\displaybreak{\crcr\noalign{\break}}}
\def\intertext@{\def\intertext##1{\crcr\noalign{%
 \penalty\postdisplaypenalty \vskip\belowdisplayskip
 \vbox{\normalbaselines\noindent##1}%
 \penalty\predisplaypenalty \vskip\abovedisplayskip}}}
\newskip\centering@
\centering@\z@ plus\@m\p@
\def\align{\relax\ifingather@\DN@{\csname align (in
  \string\gather)\endcsname}\else
 \ifmmode\ifinner\DN@{\onlydmatherr@\align}\else
  \let\next@\align@\fi
 \else\DN@{\onlydmatherr@\align}\fi\fi\next@}
\newhelp\andhelp@
{An extra & here is so disastrous that you should probably exit^^J
and fix things up.}
\newif\iftag@
\newcount\and@
\def\align@{\inalign@true\inany@true
 \vspace@\allowdisplaybreak@\displaybreak@\intertext@
 \def\tag{\global\tag@true\ifnum\and@=\z@\DN@{&&}\else
          \DN@{&}\fi\next@}%
 \iftagsleft@\DN@{\csname align \endcsname}\else
  \DN@{\csname align \space\endcsname}\fi\next@}
\def\Tag@{\iftag@\else\errhelp\andhelp@\err@{Extra & on this line}\fi}
\newdimen\lwidth@
\newdimen\rwidth@
\newdimen\maxlwidth@
\newdimen\maxrwidth@
\newdimen\totwidth@
\def\measure@#1\endalign{\lwidth@\z@\rwidth@\z@\maxlwidth@\z@\maxrwidth@\z@
 \global\and@\z@                                                            %1
 \setbox@ne\vbox                                                            %2
  {\everycr{\noalign{\global\tag@false\global\and@\z@}}\Let@                %3
  \halign{\setboxz@h{$\m@th\displaystyle{\@lign##}$}%                       %4
   \global\lwidth@\wdz@                                                     %5
   \ifdim\lwidth@>\maxlwidth@\global\maxlwidth@\lwidth@\fi                  %6
   \global\advance\and@\@ne                                                 %7
   &\setboxz@h{$\m@th\displaystyle{{}\@lign##}$}\global\rwidth@\wdz@        %8
   \ifdim\rwidth@>\maxrwidth@\global\maxrwidth@\rwidth@\fi                  %9
   \global\advance\and@\@ne                                                %10
   &\Tag@
   \eat@{##}\crcr#1\crcr}}%                                                %11
 \totwidth@\maxlwidth@\advance\totwidth@\maxrwidth@}                       %12
\def\displ@y@{\global\dt@ptrue\openup\jot
 \everycr{\noalign{\global\tag@false\global\and@\z@\ifdt@p\global\dt@pfalse
 \vskip-\lineskiplimit\vskip\normallineskiplimit\else
 \penalty\interdisplaylinepenalty\fi}}}
\def\black@#1{\noalign{\ifdim#1>\displaywidth
 \dimen@\prevdepth\nointerlineskip                                          %1
 \vskip-\ht\strutbox@\vskip-\dp\strutbox@                                   %2
 \vbox{\noindent\hbox to#1{\strut@\hfill}}%                                 %3
 \prevdepth\dimen@                                                          %4
 \fi}}
\expandafter\def\csname align \space\endcsname#1\endalign
 {\measure@#1\endalign\global\and@\z@                                       %1
 \ifingather@\everycr{\noalign{\global\and@\z@}}\else\displ@y@\fi           %2
 \Let@\tabskip\centering@                                                   %3
 \halign to\displaywidth
  {\hfil\strut@\setboxz@h{$\m@th\displaystyle{\@lign##}$}%                  %4
  \global\lwidth@\wdz@\boxz@\global\advance\and@\@ne                        %5
  \tabskip\z@skip                                                           %6
  &\setboxz@h{$\m@th\displaystyle{{}\@lign##}$}%                            %7
  \global\rwidth@\wdz@\boxz@\hfill\global\advance\and@\@ne                  %8
  \tabskip\centering@                                                       %9
  &\setboxz@h{\@lign\strut@\maketag@##\maketag@}%                          %10
  \dimen@\displaywidth\advance\dimen@-\totwidth@
  \divide\dimen@\tw@\advance\dimen@\maxrwidth@\advance\dimen@-\rwidth@     %11
  \ifdim\dimen@<\tw@\wdz@\llap{\vtop{\normalbaselines\null\boxz@}}%        %12
  \else\llap{\boxz@}\fi                                                    %13
  \tabskip\z@skip                                                          %14
  \crcr#1\crcr                                                             %15
  \black@\totwidth@}}                                                      %16
\newdimen\lineht@
\expandafter\def\csname align \endcsname#1\endalign{\measure@#1\endalign
 \global\and@\z@
 \ifdim\totwidth@>\displaywidth\let\displaywidth@\totwidth@\else
  \let\displaywidth@\displaywidth\fi                                        %1
 \ifingather@\everycr{\noalign{\global\and@\z@}}\else\displ@y@\fi
 \Let@\tabskip\centering@\halign to\displaywidth
  {\hfil\strut@\setboxz@h{$\m@th\displaystyle{\@lign##}$}%
  \global\lwidth@\wdz@\global\lineht@\ht\z@                                 %2
  \boxz@\global\advance\and@\@ne
  \tabskip\z@skip&\setboxz@h{$\m@th\displaystyle{{}\@lign##}$}%
  \global\rwidth@\wdz@\ifdim\ht\z@>\lineht@\global\lineht@\ht\z@\fi         %3
  \boxz@\hfil\global\advance\and@\@ne
  \tabskip\centering@&\kern-\displaywidth@                                  %4
  \setboxz@h{\@lign\strut@\maketag@##\maketag@}%
  \dimen@\displaywidth\advance\dimen@-\totwidth@
  \divide\dimen@\tw@\advance\dimen@\maxlwidth@\advance\dimen@-\lwidth@
  \ifdim\dimen@<\tw@\wdz@
   \rlap{\vbox{\normalbaselines\boxz@\vbox to\lineht@{}}}\else
   \rlap{\boxz@}\fi
  \tabskip\displaywidth@\crcr#1\crcr\black@\totwidth@}}
\expandafter\def\csname align (in \string\gather)\endcsname
  #1\endalign{\vcenter{\align@#1\endalign}}
\Invalid@\endalign
\newif\ifxat@
\def\alignat{\RIfMIfI@\DN@{\onlydmatherr@\alignat}\else
 \DN@{\csname alignat \endcsname}\fi\else
 \DN@{\onlydmatherr@\alignat}\fi\next@}
\newif\ifmeasuring@
\newbox\savealignat@
\expandafter\def\csname alignat \endcsname#1#2\endalignat                   %1
 {\inany@true\xat@false
 \def\tag{\global\tag@true\count@#1\relax\multiply\count@\tw@
  \xdef\tag@{}\loop\ifnum\count@>\and@\xdef\tag@{&\tag@}\advance\count@\m@ne
  \repeat\tag@}%
 \vspace@\allowdisplaybreak@\displaybreak@\intertext@
 \displ@y@\measuring@true                                                   %2
 \setbox\savealignat@\hbox{$\m@th\displaystyle\Let@
  \attag@{#1}%                                                              %3
  \vbox{\halign{\span\preamble@@\crcr#2\crcr}}$}%
 \measuring@false                                                           %4
 \Let@\attag@{#1}%                                                          %5
 \tabskip\centering@\halign to\displaywidth
  {\span\preamble@@\crcr#2\crcr                                             %6
  \black@{\wd\savealignat@}}}                                               %7
\Invalid@\endalignat
\def\xalignat{\RIfMIfI@
 \DN@{\onlydmatherr@\xalignat}\else
 \DN@{\csname xalignat \endcsname}\fi\else
 \DN@{\onlydmatherr@\xalignat}\fi\next@}
\expandafter\def\csname xalignat \endcsname#1#2\endxalignat
 {\inany@true\xat@true
 \def\tag{\global\tag@true\def\tag@{}\count@#1\relax\multiply\count@\tw@
  \loop\ifnum\count@>\and@\xdef\tag@{&\tag@}\advance\count@\m@ne\repeat\tag@}%
 \vspace@\allowdisplaybreak@\displaybreak@\intertext@
 \displ@y@\measuring@true\setbox\savealignat@\hbox{$\m@th\displaystyle\Let@
 \attag@{#1}\vbox{\halign{\span\preamble@@\crcr#2\crcr}}$}%
 \measuring@false\Let@
 \attag@{#1}\tabskip\centering@\halign to\displaywidth
 {\span\preamble@@\crcr#2\crcr\black@{\wd\savealignat@}}}
\def\attag@#1{\let\Maketag@\maketag@\let\TAG@\Tag@                          %1
 \let\Tag@=0\let\maketag@=0%                                                %2
 \ifmeasuring@\def\llap@##1{\setboxz@h{##1}\hbox to\tw@\wdz@{}}%
  \def\rlap@##1{\setboxz@h{##1}\hbox to\tw@\wdz@{}}\else
  \let\llap@\llap\let\rlap@\rlap\fi                                         %3
 \toks@{\hfil\strut@$\m@th\displaystyle{\@lign\the\hashtoks@}$\tabskip\z@skip
  \global\advance\and@\@ne&$\m@th\displaystyle{{}\@lign\the\hashtoks@}$\hfil
  \ifxat@\tabskip\centering@\fi\global\advance\and@\@ne}%                   %4
 \iftagsleft@
  \toks@@{\tabskip\centering@&\Tag@\kern-\displaywidth
   \rlap@{\@lign\maketag@\the\hashtoks@\maketag@}%
   \global\advance\and@\@ne\tabskip\displaywidth}\else
  \toks@@{\tabskip\centering@&\Tag@\llap@{\@lign\maketag@
   \the\hashtoks@\maketag@}\global\advance\and@\@ne\tabskip\z@skip}\fi      %5
 \atcount@#1\relax\advance\atcount@\m@ne
 \loop\ifnum\atcount@>\z@
 \toks@=\expandafter{\the\toks@&\hfil$\m@th\displaystyle{\@lign
  \the\hashtoks@}$\global\advance\and@\@ne
  \tabskip\z@skip&$\m@th\displaystyle{{}\@lign\the\hashtoks@}$\hfil\ifxat@
  \tabskip\centering@\fi\global\advance\and@\@ne}\advance\atcount@\m@ne
 \repeat                                                                    %6
 \xdef\preamble@{\the\toks@\the\toks@@}%                                    %7
 \xdef\preamble@@{\preamble@}%                                              %8
 \let\maketag@\Maketag@\let\Tag@\TAG@}                                      %9
\Invalid@\endxalignat
\def\xxalignat{\RIfMIfI@
 \DN@{\onlydmatherr@\xxalignat}\else\DN@{\csname xxalignat
  \endcsname}\fi\else
 \DN@{\onlydmatherr@\xxalignat}\fi\next@}
\expandafter\def\csname xxalignat \endcsname#1#2\endxxalignat{\inany@true
 \vspace@\allowdisplaybreak@\displaybreak@\intertext@
 \displ@y\setbox\savealignat@\hbox{$\m@th\displaystyle\Let@
 \xxattag@{#1}\vbox{\halign{\span\preamble@@\crcr#2\crcr}}$}%
 \Let@\xxattag@{#1}\tabskip\z@skip\halign to\displaywidth
 {\span\preamble@@\crcr#2\crcr\black@{\wd\savealignat@}}}
\def\xxattag@#1{\toks@{\tabskip\z@skip\hfil\strut@
 $\m@th\displaystyle{\the\hashtoks@}$&%
 $\m@th\displaystyle{{}\the\hashtoks@}$\hfil\tabskip\centering@&}%
 \atcount@#1\relax\advance\atcount@\m@ne\loop\ifnum\atcount@>\z@
 \toks@=\expandafter{\the\toks@&\hfil$\m@th\displaystyle{\the\hashtoks@}$%
  \tabskip\z@skip&$\m@th\displaystyle{{}\the\hashtoks@}$\hfil
  \tabskip\centering@}\advance\atcount@\m@ne\repeat
 \xdef\preamble@{\the\toks@\tabskip\z@skip}\xdef\preamble@@{\preamble@}}
\Invalid@\endxxalignat
\newdimen\gwidth@
\newdimen\gmaxwidth@
\def\gmeasure@#1\endgather{\gwidth@\z@\gmaxwidth@\z@\setbox@ne\vbox{\Let@
 \halign{\setboxz@h{$\m@th\displaystyle{##}$}\global\gwidth@\wdz@
 \ifdim\gwidth@>\gmaxwidth@\global\gmaxwidth@\gwidth@\fi
 &\eat@{##}\crcr#1\crcr}}}
\def\gather{\RIfMIfI@\DN@{\onlydmatherr@\gather}\else
 \ingather@true\inany@true\def\tag{&}%
 \vspace@\allowdisplaybreak@\displaybreak@\intertext@
 \displ@y\Let@
 \iftagsleft@\DN@{\csname gather \endcsname}\else
  \DN@{\csname gather \space\endcsname}\fi\fi
 \else\DN@{\onlydmatherr@\gather}\fi\next@}
\expandafter\def\csname gather \space\endcsname#1\endgather
 {\gmeasure@#1\endgather\tabskip\centering@
 \halign to\displaywidth{\hfil\strut@\setboxz@h{$\m@th\displaystyle{##}$}%
 \global\gwidth@\wdz@\boxz@\hfil&
 \setboxz@h{\strut@{\maketag@##\maketag@}}%
 \dimen@\displaywidth\advance\dimen@-\gwidth@
 \ifdim\dimen@>\tw@\wdz@\llap{\boxz@}\else
 \llap{\vtop{\normalbaselines\null\boxz@}}\fi
 \tabskip\z@skip\crcr#1\crcr\black@\gmaxwidth@}}
\newdimen\glineht@
\expandafter\def\csname gather \endcsname#1\endgather{\gmeasure@#1\endgather
 \ifdim\gmaxwidth@>\displaywidth\let\gdisplaywidth@\gmaxwidth@\else
 \let\gdisplaywidth@\displaywidth\fi\tabskip\centering@\halign to\displaywidth
 {\hfil\strut@\setboxz@h{$\m@th\displaystyle{##}$}%
 \global\gwidth@\wdz@\global\glineht@\ht\z@\boxz@\hfil&\kern-\gdisplaywidth@
 \setboxz@h{\strut@{\maketag@##\maketag@}}%
 \dimen@\displaywidth\advance\dimen@-\gwidth@
 \ifdim\dimen@>\tw@\wdz@\rlap{\boxz@}\else
 \rlap{\vbox{\normalbaselines\boxz@\vbox to\glineht@{}}}\fi
 \tabskip\gdisplaywidth@\crcr#1\crcr\black@\gmaxwidth@}}
\newif\ifctagsplit@
\def\CenteredTagsOnSplits{\global\ctagsplit@true}
\def\TopOrBottomTagsOnSplits{\global\ctagsplit@false}
\TopOrBottomTagsOnSplits
\def\split{\relax\ifinany@\let\next@\insplit@\else
 \ifmmode\ifinner\def\next@{\onlydmatherr@\split}\else
 \let\next@\outsplit@\fi\else
 \def\next@{\onlydmatherr@\split}\fi\fi\next@}
\def\insplit@{\global\setbox\z@\vbox\bgroup\vspace@\Let@\ialign\bgroup
 \hfil\strut@$\m@th\displaystyle{##}$&$\m@th\displaystyle{{}##}$\hfill\crcr}
\def\endsplit{\crcr\egroup\egroup\iftagsleft@\expandafter\lendsplit@\else
 \expandafter\rendsplit@\fi}
\def\rendsplit@{\global\setbox9 \vbox
 {\unvcopy\z@\global\setbox8 \lastbox\unskip}%                              %1
 \setbox@ne\hbox{\unhcopy8 \unskip\global\setbox\tw@\lastbox
 \unskip\global\setbox\thr@@\lastbox}%                                      %2
 \global\setbox7 \hbox{\unhbox\tw@\unskip}%                                 %3
 \ifinalign@\ifctagsplit@                                                   %4
  \gdef\split@{\hbox to\wd\thr@@{}&
   \vcenter{\vbox{\moveleft\wd\thr@@\boxz@}}}%                              %5
 \else\gdef\split@{&\vbox{\moveleft\wd\thr@@\box9}\crcr
  \box\thr@@&\box7}\fi                                                      %6
 \else                                                                      %7
  \ifctagsplit@\gdef\split@{\vcenter{\boxz@}}\else
  \gdef\split@{\box9\crcr\hbox{\box\thr@@\box7}}\fi
 \fi
 \split@}                                                                   %8
\def\lendsplit@{\global\setbox9\vtop{\unvcopy\z@}%                          %1
 \setbox@ne\vbox{\unvcopy\z@\global\setbox8\lastbox}%                       %2
 \setbox@ne\hbox{\unhcopy8\unskip\setbox\tw@\lastbox
  \unskip\global\setbox\thr@@\lastbox}%                                     %3
 \ifinalign@\ifctagsplit@                                                   %4
  \gdef\split@{\hbox to\wd\thr@@{}&
  \vcenter{\vbox{\moveleft\wd\thr@@\box9}}}%                                %5
  \else                                                                     %6
  \gdef\split@{\hbox to\wd\thr@@{}&\vbox{\moveleft\wd\thr@@\box9}}\fi
 \else
  \ifctagsplit@\gdef\split@{\vcenter{\box9}}\else
  \gdef\split@{\box9}\fi
 \fi\split@}
\def\outsplit@#1$${\align\insplit@#1\endalign$$}
\newdimen\multlinegap@
\multlinegap@1em
\newdimen\multlinetaggap@
\multlinetaggap@1em
\def\MultlineGap#1{\global\multlinegap@#1\relax}
\def\multlinegap#1{\RIfMIfI@\onlydmatherr@\multlinegap\else
 \multlinegap@#1\relax\fi\else\onlydmatherr@\multlinegap\fi}
\def\nomultlinegap{\multlinegap{\z@}}
\def\multline{\RIfMIfI@
 \DN@{\onlydmatherr@\multline}\else
 \DN@{\multline@}\fi\else
 \DN@{\onlydmatherr@\multline}\fi\next@}
\newif\iftagin@
\def\tagin@#1{\tagin@false\in@\tag{#1}\ifin@\tagin@true\fi}
\def\multline@#1$${\inany@true\vspace@\allowdisplaybreak@\displaybreak@
 \tagin@{#1}\iftagsleft@\DN@{\multline@l#1$$}\else
 \DN@{\multline@r#1$$}\fi\next@}
\newdimen\mwidth@
\def\rmmeasure@#1\endmultline{%
 \def\shoveleft##1{##1}\def\shoveright##1{##1}%                             %1
 \setbox@ne\vbox{\Let@\halign{\setboxz@h
  {$\m@th\@lign\displaystyle{}##$}\global\mwidth@\wdz@
  \crcr#1\crcr}}}
\newdimen\mlineht@
\newif\ifzerocr@
\newif\ifonecr@
\def\lmmeasure@#1\endmultline{\global\zerocr@true\global\onecr@false
 \everycr{\noalign{\ifonecr@\global\onecr@false\fi
  \ifzerocr@\global\zerocr@false\global\onecr@true\fi}}%                    %1
  \def\shoveleft##1{##1}\def\shoveright##1{##1}%
 \setbox@ne\vbox{\Let@\halign{\setboxz@h
  {$\m@th\@lign\displaystyle{}##$}\ifonecr@\global\mwidth@\wdz@
  \global\mlineht@\ht\z@\fi\crcr#1\crcr}}}
\newbox\mtagbox@
\newdimen\ltwidth@
\newdimen\rtwidth@
\def\multline@l#1$${\iftagin@\DN@{\lmultline@@#1$$}\else
 \DN@{\setbox\mtagbox@\null\ltwidth@\z@\rtwidth@\z@
  \lmultline@@@#1$$}\fi\next@}
\def\lmultline@@#1\endmultline\tag#2$${%
 \setbox\mtagbox@\hbox{\maketag@#2\maketag@}%                               %1
 \lmmeasure@#1\endmultline\dimen@\mwidth@\advance\dimen@\wd\mtagbox@
 \advance\dimen@\multlinetaggap@                                            %2
 \ifdim\dimen@>\displaywidth\ltwidth@\z@\else\ltwidth@\wd\mtagbox@\fi       %3
 \lmultline@@@#1\endmultline$$}
\def\lmultline@@@{\displ@y
 \def\shoveright##1{##1\hfilneg\hskip\multlinegap@}%
 \def\shoveleft##1{\setboxz@h{$\m@th\displaystyle{}##1$}%
  \setbox@ne\hbox{$\m@th\displaystyle##1$}%
  \hfilneg
  \iftagin@
   \ifdim\ltwidth@>\z@\hskip\ltwidth@\hskip\multlinetaggap@\fi
  \else\hskip\multlinegap@\fi\hskip.5\wd@ne\hskip-.5\wdz@##1}%              %1
  \halign\bgroup\Let@\hbox to\displaywidth
   {\strut@$\m@th\displaystyle\hfil{}##\hfil$}\crcr
   \hfilneg                                                                 %2
   \iftagin@                                                                %3
    \ifdim\ltwidth@>\z@                                                     %4
     \box\mtagbox@\hskip\multlinetaggap@                                    %5
    \else
     \rlap{\vbox{\normalbaselines\hbox{\strut@\box\mtagbox@}%
     \vbox to\mlineht@{}}}\fi                                               %6
   \else\hskip\multlinegap@\fi}                                             %7
\def\multline@r#1$${\iftagin@\DN@{\rmultline@@#1$$}\else
 \DN@{\setbox\mtagbox@\null\ltwidth@\z@\rtwidth@\z@
  \rmultline@@@#1$$}\fi\next@}
\def\rmultline@@#1\endmultline\tag#2$${\ltwidth@\z@
 \setbox\mtagbox@\hbox{\maketag@#2\maketag@}%
 \rmmeasure@#1\endmultline\dimen@\mwidth@\advance\dimen@\wd\mtagbox@
 \advance\dimen@\multlinetaggap@
 \ifdim\dimen@>\displaywidth\rtwidth@\z@\else\rtwidth@\wd\mtagbox@\fi
 \rmultline@@@#1\endmultline$$}
\def\rmultline@@@{\displ@y
 \def\shoveright##1{##1\hfilneg\iftagin@\ifdim\rtwidth@>\z@
  \hskip\rtwidth@\hskip\multlinetaggap@\fi\else\hskip\multlinegap@\fi}%
 \def\shoveleft##1{\setboxz@h{$\m@th\displaystyle{}##1$}%
  \setbox@ne\hbox{$\m@th\displaystyle##1$}%
  \hfilneg\hskip\multlinegap@\hskip.5\wd@ne\hskip-.5\wdz@##1}%
 \halign\bgroup\Let@\hbox to\displaywidth
  {\strut@$\m@th\displaystyle\hfil{}##\hfil$}\crcr
 \hfilneg\hskip\multlinegap@}
\def\endmultline{\iftagsleft@\expandafter\lendmultline@\else
 \expandafter\rendmultline@\fi}
\def\lendmultline@{\hfilneg\hskip\multlinegap@\crcr\egroup}
\def\rendmultline@{\iftagin@                                                %1
 \ifdim\rtwidth@>\z@                                                        %2
  \hskip\multlinetaggap@\box\mtagbox@                                       %3
 \else\llap{\vtop{\normalbaselines\null\hbox{\strut@\box\mtagbox@}}}\fi     %4
 \else\hskip\multlinegap@\fi                                                %5
 \hfilneg\crcr\egroup}
\def\bmod{\mskip-\medmuskip\mkern5mu\mathbin{\fam\z@ mod}\penalty900
 \mkern5mu\mskip-\medmuskip}
\def\pmod#1{\allowbreak\ifinner\mkern8mu\else\mkern18mu\fi
 ({\fam\z@ mod}\,\,#1)}
\def\pod#1{\allowbreak\ifinner\mkern8mu\else\mkern18mu\fi(#1)}
\def\mod#1{\allowbreak\ifinner\mkern12mu\else\mkern18mu\fi{\fam\z@ mod}\,\,#1}
\message{continued fractions,}
\newcount\cfraccount@
\def\cfrac{\bgroup\bgroup\advance\cfraccount@\@ne\strut
 \iffalse{\fi\def\\{\over\displaystyle}\iffalse}\fi}
\def\lcfrac{\bgroup\bgroup\advance\cfraccount@\@ne\strut
 \iffalse{\fi\def\\{\hfill\over\displaystyle}\iffalse}\fi}
\def\rcfrac{\bgroup\bgroup\advance\cfraccount@\@ne\strut\hfill
 \iffalse{\fi\def\\{\over\displaystyle}\iffalse}\fi}
\def\gloop@#1\repeat{\gdef\body{#1}\iterate}
\def\endcfrac{\gloop@\ifnum\cfraccount@>\z@\global\advance\cfraccount@\m@ne
 \egroup\hskip-\nulldelimiterspace\egroup\repeat}
\message{compound symbols,}
\def\binrel@#1{\setboxz@h{\thinmuskip0mu
  \medmuskip\m@ne mu\thickmuskip\@ne mu$#1\m@th$}%
 \setbox@ne\hbox{\thinmuskip0mu\medmuskip\m@ne mu\thickmuskip
  \@ne mu${}#1{}\m@th$}%
 \setbox\tw@\hbox{\hskip\wd@ne\hskip-\wdz@}}
\def\overset#1\to#2{\binrel@{#2}\ifdim\wd\tw@<\z@
 \mathbin{\mathop{\kern\z@#2}\limits^{#1}}\else\ifdim\wd\tw@>\z@
 \mathrel{\mathop{\kern\z@#2}\limits^{#1}}\else
 {\mathop{\kern\z@#2}\limits^{#1}}{}\fi\fi}
\def\underset#1\to#2{\binrel@{#2}\ifdim\wd\tw@<\z@
 \mathbin{\mathop{\kern\z@#2}\limits_{#1}}\else\ifdim\wd\tw@>\z@
 \mathrel{\mathop{\kern\z@#2}\limits_{#1}}\else
 {\mathop{\kern\z@#2}\limits_{#1}}{}\fi\fi}
\def\oversetbrace#1\to#2{\overbrace{#2}^{#1}}
\def\undersetbrace#1\to#2{\underbrace{#2}_{#1}}
\def\sideset#1\and#2\to#3{%
 \setbox@ne\hbox{$\dsize{\vphantom{#3}}#1{#3}\m@th$}%
 \setbox\tw@\hbox{$\dsize{#3}#2\m@th$}%
 \hskip\wd@ne\hskip-\wd\tw@\mathop{\hskip\wd\tw@\hskip-\wd@ne
  {\vphantom{#3}}#1{#3}#2}}
\def\rightarrowfill@#1{\setboxz@h{$#1-\m@th$}\ht\z@\z@
  $#1\m@th\copy\z@\mkern-6mu\cleaders
  \hbox{$#1\mkern-2mu\box\z@\mkern-2mu$}\hfill
  \mkern-6mu\mathord\rightarrow$}
\def\leftarrowfill@#1{\setboxz@h{$#1-\m@th$}\ht\z@\z@
  $#1\m@th\mathord\leftarrow\mkern-6mu\cleaders
  \hbox{$#1\mkern-2mu\copy\z@\mkern-2mu$}\hfill
  \mkern-6mu\box\z@$}
\def\leftrightarrowfill@#1{\setboxz@h{$#1-\m@th$}\ht\z@\z@
  $#1\m@th\mathord\leftarrow\mkern-6mu\cleaders
  \hbox{$#1\mkern-2mu\box\z@\mkern-2mu$}\hfill
  \mkern-6mu\mathord\rightarrow$}
\def\overrightarrow{\mathpalette\overrightarrow@}
\def\overrightarrow@#1#2{\vbox{\ialign{##\crcr\rightarrowfill@#1\crcr
 \noalign{\kern-\ex@\nointerlineskip}$\m@th\hfil#1#2\hfil$\crcr}}}

\def\overleftarrow{\mathpalette\overleftarrow@}
\def\overleftarrow@#1#2{\vbox{\ialign{##\crcr\leftarrowfill@#1\crcr
 \noalign{\kern-\ex@\nointerlineskip}$\m@th\hfil#1#2\hfil$\crcr}}}
\def\overleftrightarrow{\mathpalette\overleftrightarrow@}
\def\overleftrightarrow@#1#2{\vbox{\ialign{##\crcr\leftrightarrowfill@#1\crcr
 \noalign{\kern-\ex@\nointerlineskip}$\m@th\hfil#1#2\hfil$\crcr}}}
\def\underrightarrow{\mathpalette\underrightarrow@}
\def\underrightarrow@#1#2{\vtop{\ialign{##\crcr$\m@th\hfil#1#2\hfil$\crcr
 \noalign{\nointerlineskip}\rightarrowfill@#1\crcr}}}

\def\underleftarrow{\mathpalette\underleftarrow@}
\def\underleftarrow@#1#2{\vtop{\ialign{##\crcr$\m@th\hfil#1#2\hfil$\crcr
 \noalign{\nointerlineskip}\leftarrowfill@#1\crcr}}}
\def\underleftrightarrow{\mathpalette\underleftrightarrow@}
\def\underleftrightarrow@#1#2{\vtop{\ialign{##\crcr$\m@th\hfil#1#2\hfil$\crcr
 \noalign{\nointerlineskip}\leftrightarrowfill@#1\crcr}}}
\message{various kinds of dots,}
\let\DOTSI\relax
\let\DOTSB\relax

\newif\ifmath@
{\uccode`7=`\\ \uccode`8=`m \uccode`9=`a \uccode`0=`t \uccode`!=`h
 \uppercase{\gdef\math@#1#2#3#4#5#6\math@{\global\math@false\ifx 7#1\ifx 8#2%
 \ifx 9#3\ifx 0#4\ifx !#5\xdef\meaning@{#6}\global\math@true\fi\fi\fi\fi\fi}}}
\newif\ifmathch@
{\uccode`7=`c \uccode`8=`h \uccode`9=`\"
 \uppercase{\gdef\mathch@#1#2#3#4#5#6\mathch@{\global\mathch@false
  \ifx 7#1\ifx 8#2\ifx 9#5\global\mathch@true\xdef\meaning@{9#6}\fi\fi\fi}}}
\newcount\classnum@
\def\getmathch@#1.#2\getmathch@{\classnum@#1 \divide\classnum@4096
 \ifcase\number\classnum@\or\or\gdef\thedots@{\dotsb@}\or
 \gdef\thedots@{\dotsb@}\fi}
\newif\ifmathbin@
{\uccode`4=`b \uccode`5=`i \uccode`6=`n
 \uppercase{\gdef\mathbin@#1#2#3{\relaxnext@
  \DNii@##1\mathbin@{\ifx\space@\next\global\mathbin@true\fi}%
 \global\mathbin@false\DN@##1\mathbin@{}%
 \ifx 4#1\ifx 5#2\ifx 6#3\DN@{\FN@\nextii@}\fi\fi\fi\next@}}}
\newif\ifmathrel@
{\uccode`4=`r \uccode`5=`e \uccode`6=`l
 \uppercase{\gdef\mathrel@#1#2#3{\relaxnext@
  \DNii@##1\mathrel@{\ifx\space@\next\global\mathrel@true\fi}%
 \global\mathrel@false\DN@##1\mathrel@{}%
 \ifx 4#1\ifx 5#2\ifx 6#3\DN@{\FN@\nextii@}\fi\fi\fi\next@}}}
\newif\ifmacro@
{\uccode`5=`m \uccode`6=`a \uccode`7=`c
 \uppercase{\gdef\macro@#1#2#3#4\macro@{\global\macro@false
  \ifx 5#1\ifx 6#2\ifx 7#3\global\macro@true
  \xdef\meaning@{\macro@@#4\macro@@}\fi\fi\fi}}}
\def\macro@@#1->#2\macro@@{#2}
\newif\ifDOTS@
\newcount\DOTSCASE@
{\uccode`6=`\\ \uccode`7=`D \uccode`8=`O \uccode`9=`T \uccode`0=`S
 \uppercase{\gdef\DOTS@#1#2#3#4#5{\global\DOTS@false\DN@##1\DOTS@{}%
  \ifx 6#1\ifx 7#2\ifx 8#3\ifx 9#4\ifx 0#5\let\next@\DOTS@@\fi\fi\fi\fi\fi
  \next@}}}
{\uccode`3=`B \uccode`4=`I \uccode`5=`X
 \uppercase{\gdef\DOTS@@#1{\relaxnext@
  \DNii@##1\DOTS@{\ifx\space@\next\global\DOTS@true\fi}%
  \DN@{\FN@\nextii@}%
  \ifx 3#1\global\DOTSCASE@\z@\else
  \ifx 4#1\global\DOTSCASE@\@ne\else
  \ifx 5#1\global\DOTSCASE@\tw@\else\DN@##1\DOTS@{}%
  \fi\fi\fi\next@}}}
\newif\ifnot@
{\uccode`5=`\\ \uccode`6=`n \uccode`7=`o \uccode`8=`t
 \uppercase{\gdef\not@#1#2#3#4{\relaxnext@
  \DNii@##1\not@{\ifx\space@\next\global\not@true\fi}%
 \global\not@false\DN@##1\not@{}%
 \ifx 5#1\ifx 6#2\ifx 7#3\ifx 8#4\DN@{\FN@\nextii@}\fi\fi\fi
 \fi\next@}}}
\newif\ifkeybin@
\def\keybin@{\keybin@true
 \ifx\next+\else\ifx\next=\else\ifx\next<\else\ifx\next>\else\ifx\next-\else
 \ifx\next*\else\ifx\next:\else\keybin@false\fi\fi\fi\fi\fi\fi\fi}
\def\dots{\RIfM@\expandafter\mdots@\else\expandafter\tdots@\fi}
\def\tdots@{\unskip\relaxnext@
 \DN@{$\m@th\mathinner{\ldotp\ldotp\ldotp}\,
   \ifx\next,\,$\else\ifx\next.\,$\else\ifx\next;\,$\else\ifx\next:\,$\else
   \ifx\next?\,$\else\ifx\next!\,$\else$ \fi\fi\fi\fi\fi\fi}%
 \ \FN@\next@}
\def\mdots@{\FN@\mdots@@}
\def\mdots@@{\gdef\thedots@{\dotso@}%                                       %1
 \ifx\next\boldkey\gdef\thedots@\boldkey{\boldkeydots@}\else                %2
 \ifx\next\boldsymbol\gdef\thedots@\boldsymbol{\boldsymboldots@}\else       %3
 \ifx,\next\gdef\thedots@{\dotsc}%                                          %4
 \else\ifx\not\next\gdef\thedots@{\dotsb@}%                                 %5
 \else\keybin@
 \ifkeybin@\gdef\thedots@{\dotsb@}%                                         %6
 \else\xdef\meaning@{\meaning\next..........}\xdef\meaning@@{\meaning@}%    %7
  \expandafter\math@\meaning@\math@
  \ifmath@
   \expandafter\mathch@\meaning@\mathch@
   \ifmathch@\expandafter\getmathch@\meaning@\getmathch@\fi                 %8
  \else\expandafter\macro@\meaning@@\macro@                                 %9
  \ifmacro@                                                                %10
   \expandafter\not@\meaning@\not@\ifnot@\gdef\thedots@{\dotsb@}%          %11
  \else\expandafter\DOTS@\meaning@\DOTS@
  \ifDOTS@
   \ifcase\number\DOTSCASE@\gdef\thedots@{\dotsb@}%
    \or\gdef\thedots@{\dotsi}\else\fi                                      %12
  \else\expandafter\math@\meaning@\math@                                   %13
  \ifmath@\expandafter\mathbin@\meaning@\mathbin@
  \ifmathbin@\gdef\thedots@{\dotsb@}%                                      %14
  \else\expandafter\mathrel@\meaning@\mathrel@
  \ifmathrel@\gdef\thedots@{\dotsb@}%                                      %15
  \fi\fi\fi\fi\fi\fi\fi\fi\fi\fi\fi\fi
 \thedots@}
\def\plainldots@{\mathinner{\ldotp\ldotp\ldotp}}
\def\plaincdots@{\mathinner{\cdotp\cdotp\cdotp}}
\def\dotsi{\!\plaincdots@}
\let\dotsb@\plaincdots@
\newif\ifextra@
\newif\ifrightdelim@
\def\rightdelim@{\global\rightdelim@true                                    %1
 \ifx\next)\else                                                            %2
 \ifx\next]\else
 \ifx\next\rbrack\else
 \ifx\next\}\else
 \ifx\next\rbrace\else
 \ifx\next\rangle\else
 \ifx\next\rceil\else
 \ifx\next\rfloor\else
 \ifx\next\rgroup\else
 \ifx\next\rmoustache\else
 \ifx\next\right\else
 \ifx\next\bigr\else
 \ifx\next\biggr\else
 \ifx\next\Bigr\else                                                        %3
 \ifx\next\Biggr\else\global\rightdelim@false
 \fi\fi\fi\fi\fi\fi\fi\fi\fi\fi\fi\fi\fi\fi\fi}
\def\extra@{%
 \global\extra@false\rightdelim@\ifrightdelim@\global\extra@true            %1
 \else\ifx\next \global\extra@true                                          %2
 \else\xdef\meaning@{\meaning\next..........}%                              %3
 \expandafter\macro@\meaning@\macro@\ifmacro@                               %4
 \expandafter\DOTS@\meaning@\DOTS@
 \ifDOTS@
 \ifnum\DOTSCASE@=\tw@\global\extra@true                                    %5
 \fi\fi\fi\fi\fi}
\newif\ifbold@
\def\dotso@{\relaxnext@
 \ifbold@
  \let\next\delayed@
  \DNii@{\extra@\plainldots@\ifextra@\,\fi}%
 \else
  \DNii@{\DN@{\extra@\plainldots@\ifextra@\,\fi}\FN@\next@}%
 \fi
 \nextii@}
\def\extrap@#1{%
 \ifx\next,\DN@{#1\,}\else
 \ifx\next;\DN@{#1\,}\else
 \ifx\next.\DN@{#1\,}\else\extra@
 \ifextra@\DN@{#1\,}\else
 \let\next@#1\fi\fi\fi\fi\next@}
\def\ldots{\DN@{\extrap@\plainldots@}%
 \FN@\next@}
\def\cdots{\DN@{\extrap@\plaincdots@}%
 \FN@\next@}

\def\dotsc{\relaxnext@
 \DN@{\ifx\next;\plainldots@\,\else
  \ifx\next.\plainldots@\,\else\extra@\plainldots@
  \ifextra@\,\fi\fi\fi}%
 \FN@\next@}
\def\cdot{\mathchar"2201 }

\message{special superscripts,}
\def\dddot#1{{\mathop{#1}\limits^{\vbox to-1.4\ex@{\kern-\tw@\ex@
 \hbox{\rm...}\vss}}}}
\def\ddddot#1{{\mathop{#1}\limits^{\vbox to-1.4\ex@{\kern-\tw@\ex@
 \hbox{\rm....}\vss}}}}
\def\sphat{^{\mathchoice{}{}%
 {\,\,\botsmash{\hbox{\lower4\ex@\hbox{$\m@th\widehat{\null}$}}}}%
 {\,\botsmash{\hbox{\lower3\ex@\hbox{$\m@th\hat{\null}$}}}}}}

\def\spacute{^{\!\botsmash{\hbox{\lower\@ne ex\hbox{\'{}}}}}}
\def\spgrave{^{\mathchoice{}{}{}{\!}%
 \botsmash{\hbox{\lower\@ne ex\hbox{\`{}}}}}}
\def\spdot{^{\hbox{\raise\ex@\hbox{\rm.}}}}
\def\spddot{^{\hbox{\raise\ex@\hbox{\rm..}}}}
\def\spdddot{^{\hbox{\raise\ex@\hbox{\rm...}}}}
\def\spddddot{^{\hbox{\raise\ex@\hbox{\rm....}}}}
\def\spbreve{^{\!\botsmash{\hbox{\lower4\ex@\hbox{\u{}}}}}}

\message{\string\text,}
\def\textonlyfont@#1#2{\def#1{\RIfM@
 \Err@{Use \string#1\space only in text}\else#2\fi}}
\textonlyfont@\rm\tenrm
\textonlyfont@\it\tenit
\textonlyfont@\sl\tensl
\textonlyfont@\bf\tenbf
\def\oldnos#1{\RIfM@{\mathcode`\,="013B \fam\@ne#1}\else
 \leavevmode\hbox{$\m@th\mathcode`\,="013B \fam\@ne#1$}\fi}
\def\text{\RIfM@\expandafter\text@\else\expandafter\text@@\fi}
\def\text@@#1{\leavevmode\hbox{#1}}
\def\mathhexbox@#1#2#3{\text{$\m@th\mathchar"#1#2#3$}}
\def\dag{{\mathhexbox@279}}
\def\ddag{{\mathhexbox@27A}}
\def\S{{\mathhexbox@278}}
\def\P{{\mathhexbox@27B}}
\newif\iffirstchoice@
\firstchoice@true
\def\text@#1{\mathchoice
 {\hbox{\everymath{\displaystyle}\def\textfonti{\the\textfont\@ne}%
  \def\textfontii{\the\textfont\tw@}\textdef@@ T#1}}
 {\hbox{\firstchoice@false
  \everymath{\textstyle}\def\textfonti{\the\textfont\@ne}%
  \def\textfontii{\the\textfont\tw@}\textdef@@ T#1}}
 {\hbox{\firstchoice@false
  \everymath{\scriptstyle}\def\textfonti{\the\scriptfont\@ne}%
  \def\textfontii{\the\scriptfont\tw@}\textdef@@ S\rm#1}}
 {\hbox{\firstchoice@false
  \everymath{\scriptscriptstyle}\def\textfonti
  {\the\scriptscriptfont\@ne}%
  \def\textfontii{\the\scriptscriptfont\tw@}\textdef@@ s\rm#1}}}
\def\textdef@@#1{\textdef@#1\rm\textdef@#1\bf\textdef@#1\sl\textdef@#1\it}
\def\rmfam{0}
\def\textdef@#1#2{%
 \DN@{\csname\expandafter\eat@\string#2fam\endcsname}%
 \if S#1\edef#2{\the\scriptfont\next@\relax}%
 \else\if s#1\edef#2{\the\scriptscriptfont\next@\relax}%
 \else\edef#2{\the\textfont\next@\relax}\fi\fi}
\scriptfont\itfam\tenit \scriptscriptfont\itfam\tenit
\scriptfont\slfam\tensl \scriptscriptfont\slfam\tensl
\newif\iftopfolded@
\newif\ifbotfolded@
\def\topfoldedtext{\topfolded@true\botfolded@false\foldedtext@}
\def\botfoldedtext{\botfolded@true\topfolded@false\foldedtext@}
\def\foldedtext{\topfolded@false\botfolded@false\foldedtext@}
\Invalid@\foldedwidth
\def\foldedtext@{\relaxnext@
 \DN@{\ifx\next\foldedwidth\let\next@\nextii@\else
  \DN@{\nextii@\foldedwidth{.3\hsize}}\fi\next@}%
 \DNii@\foldedwidth##1##2{\setbox\z@\vbox
  {\normalbaselines\hsize##1\relax
  \tolerance1600 \noindent\ignorespaces##2}\ifbotfolded@\boxz@\else
  \iftopfolded@\vtop{\unvbox\z@}\else\vcenter{\boxz@}\fi\fi}%
 \FN@\next@}
\message{math font commands,}
\def\bold{\RIfM@\expandafter\bold@\else
 \expandafter\nonmatherr@\expandafter\bold\fi}
\def\bold@#1{{\bold@@{#1}}}
\def\bold@@#1{\fam\bffam\relax#1}
\def\slanted{\RIfM@\expandafter\slanted@\else
 \expandafter\nonmatherr@\expandafter\slanted\fi}
\def\slanted@#1{{\slanted@@{#1}}}
\def\slanted@@#1{\fam\slfam\relax#1}
\def\roman{\RIfM@\expandafter\roman@\else
 \expandafter\nonmatherr@\expandafter\roman\fi}
\def\roman@#1{{\roman@@{#1}}}
\def\roman@@#1{\fam\rmfam\relax#1}
\def\italic{\RIfM@\expandafter\italic@\else
 \expandafter\nonmatherr@\expandafter\italic\fi}
\def\italic@#1{{\italic@@{#1}}}
\def\italic@@#1{\fam\itfam\relax#1}
\def\Cal{\RIfM@\expandafter\Cal@\else
 \expandafter\nonmatherr@\expandafter\Cal\fi}
\def\Cal@#1{{\Cal@@{#1}}}
\def\Cal@@#1{\noaccents@\fam\tw@#1}
\mathchardef\Gamma="0000
\mathchardef\Delta="0001
\mathchardef\Theta="0002
\mathchardef\Lambda="0003
\mathchardef\Xi="0004
\mathchardef\Pi="0005
\mathchardef\Sigma="0006
\mathchardef\Upsilon="0007
\mathchardef\Phi="0008
\mathchardef\Psi="0009
\mathchardef\Omega="000A
\mathchardef\varGamma="0100
\mathchardef\varDelta="0101
\mathchardef\varTheta="0102
\mathchardef\varLambda="0103
\mathchardef\varXi="0104
\mathchardef\varPi="0105
\mathchardef\varSigma="0106
\mathchardef\varUpsilon="0107
\mathchardef\varPhi="0108
\mathchardef\varPsi="0109
\mathchardef\varOmega="010A
\let\alloc@@\alloc@
\def\hexnumber@#1{\ifcase#1 0\or 1\or 2\or 3\or 4\or 5\or 6\or 7\or 8\or
 9\or A\or B\or C\or D\or E\or F\fi}
\def\loadmsam{%
 \font@\tenmsa=msam10
 \font@\sevenmsa=msam7
 \font@\fivemsa=msam5
 \alloc@@8\fam\chardef\sixt@@n\msafam
 \textfont\msafam=\tenmsa
 \scriptfont\msafam=\sevenmsa
 \scriptscriptfont\msafam=\fivemsa
 \edef\next{\hexnumber@\msafam}%
 \mathchardef\dabar@"0\next39
 \edef\dashrightarrow{\mathrel{\dabar@\dabar@\mathchar"0\next4B}}%
 \edef\dashleftarrow{\mathrel{\mathchar"0\next4C\dabar@\dabar@}}%
 \let\dasharrow\dashrightarrow
 \edef\ulcorner{\delimiter"4\next70\next70 }%
 \edef\urcorner{\delimiter"5\next71\next71 }%
 \edef\llcorner{\delimiter"4\next78\next78 }%
 \edef\lrcorner{\delimiter"5\next79\next79 }%
 \edef\yen{{\noexpand\mathhexbox@\next55}}%
 \edef\checkmark{{\noexpand\mathhexbox@\next58}}%
 \edef\circledR{{\noexpand\mathhexbox@\next72}}%
 \edef\maltese{{\noexpand\mathhexbox@\next7A}}%
 \global\let\loadmsam\empty}%
\def\loadmsbm{%
 \font@\tenmsb=msbm10 \font@\sevenmsb=msbm7 \font@\fivemsb=msbm5
 \alloc@@8\fam\chardef\sixt@@n\msbfam
 \textfont\msbfam=\tenmsb
 \scriptfont\msbfam=\sevenmsb \scriptscriptfont\msbfam=\fivemsb
 \global\let\loadmsbm\empty
 }
\def\widehat#1{\ifx\undefined\msbfam \DN@{362}%
  \else \setboxz@h{$\m@th#1$}%
    \edef\next@{\ifdim\wdz@>\tw@ em%
        \hexnumber@\msbfam 5B%
      \else 362\fi}\fi
  \mathaccent"0\next@{#1}}
\def\widetilde#1{\ifx\undefined\msbfam \DN@{365}%
  \else \setboxz@h{$\m@th#1$}%
    \edef\next@{\ifdim\wdz@>\tw@ em%
        \hexnumber@\msbfam 5D%
      \else 365\fi}\fi
  \mathaccent"0\next@{#1}}
\message{\string\newsymbol,}
\def\newsymbol#1#2#3#4#5{\define#1{}%
  \count@#2\relax \advance\count@\m@ne % to push case 0 to the \else clause
 \ifcase\count@
   \ifx\undefined\msafam\loadmsam\fi \let\next@\msafam
 \or \ifx\undefined\msbfam\loadmsbm\fi \let\next@\msbfam
 \else  \Err@{\Invalid@@\string\newsymbol}\let\next@\tw@\fi
 \mathchardef#1="#3\hexnumber@\next@#4#5\space}

\def\Bbb{\RIfM@\expandafter\Bbb@\else
 \expandafter\nonmatherr@\expandafter\Bbb\fi}
\def\Bbb@#1{{\Bbb@@{#1}}}
\def\Bbb@@#1{\noaccents@\fam\msbfam\relax#1}
\message{bold Greek and bold symbols,}
\def\loadbold{%
 \font@\tencmmib=cmmib10 \font@\sevencmmib=cmmib7 \font@\fivecmmib=cmmib5
 \skewchar\tencmmib'177 \skewchar\sevencmmib'177 \skewchar\fivecmmib'177
 \alloc@@8\fam\chardef\sixt@@n\cmmibfam
 \textfont\cmmibfam\tencmmib
 \scriptfont\cmmibfam\sevencmmib \scriptscriptfont\cmmibfam\fivecmmib
 \font@\tencmbsy=cmbsy10 \font@\sevencmbsy=cmbsy7 \font@\fivecmbsy=cmbsy5
 \skewchar\tencmbsy'60 \skewchar\sevencmbsy'60 \skewchar\fivecmbsy'60
 \alloc@@8\fam\chardef\sixt@@n\cmbsyfam
 \textfont\cmbsyfam\tencmbsy
 \scriptfont\cmbsyfam\sevencmbsy \scriptscriptfont\cmbsyfam\fivecmbsy
 \let\loadbold\empty
}
\def\boldnotloaded#1{\Err@{\ifcase#1\or First\else Second\fi
       bold symbol font not loaded}}
\def\mathchari@#1#2#3{\ifx\undefined\cmmibfam
    \boldnotloaded@\@ne
  \else\mathchar"#1\hexnumber@\cmmibfam#2#3\space \fi}
\def\mathcharii@#1#2#3{\ifx\undefined\cmbsyfam
    \boldnotloaded\tw@
  \else \mathchar"#1\hexnumber@\cmbsyfam#2#3\space\fi}
\edef\bffam@{\hexnumber@\bffam}
\def\boldkey#1{\ifcat\noexpand#1A%
  \ifx\undefined\cmmibfam \boldnotloaded\@ne
  \else {\fam\cmmibfam#1}\fi
 \else
 \ifx#1!\mathchar"5\bffam@21 \else
 \ifx#1(\mathchar"4\bffam@28 \else\ifx#1)\mathchar"5\bffam@29 \else
 \ifx#1+\mathchar"2\bffam@2B \else\ifx#1:\mathchar"3\bffam@3A \else
 \ifx#1;\mathchar"6\bffam@3B \else\ifx#1=\mathchar"3\bffam@3D \else
 \ifx#1?\mathchar"5\bffam@3F \else\ifx#1[\mathchar"4\bffam@5B \else
 \ifx#1]\mathchar"5\bffam@5D \else
 \ifx#1,\mathchari@63B \else
 \ifx#1-\mathcharii@200 \else
 \ifx#1.\mathchari@03A \else
 \ifx#1/\mathchari@03D \else
 \ifx#1<\mathchari@33C \else
 \ifx#1>\mathchari@33E \else
 \ifx#1*\mathcharii@203 \else
 \ifx#1|\mathcharii@06A \else
 \ifx#10\bold0\else\ifx#11\bold1\else\ifx#12\bold2\else\ifx#13\bold3\else
 \ifx#14\bold4\else\ifx#15\bold5\else\ifx#16\bold6\else\ifx#17\bold7\else
 \ifx#18\bold8\else\ifx#19\bold9\else
  \Err@{\string\boldkey\space can't be used with #1}%
 \fi\fi\fi\fi\fi\fi\fi\fi\fi\fi\fi\fi\fi\fi\fi
 \fi\fi\fi\fi\fi\fi\fi\fi\fi\fi\fi\fi\fi\fi}
\def\boldsymbol#1{%
 \DN@{\Err@{You can't use \string\boldsymbol\space with \string#1}#1}%
 \ifcat\noexpand#1A%
   \let\next@\relax
   \ifx\undefined\cmmibfam \boldnotloaded\@ne
   \else {\fam\cmmibfam#1}\fi
 \else
  \xdef\meaning@{\meaning#1.........}%
  \expandafter\math@\meaning@\math@
  \ifmath@
   \expandafter\mathch@\meaning@\mathch@
   \ifmathch@
    \expandafter\boldsymbol@@\meaning@\boldsymbol@@
   \fi
  \else
   \expandafter\macro@\meaning@\macro@
   \expandafter\delim@\meaning@\delim@
   \ifdelim@
    \expandafter\delim@@\meaning@\delim@@
   \else
    \boldsymbol@{#1}%
   \fi
  \fi
 \fi
 \next@}
\def\mathhexboxii@#1#2{\ifx\undefined\cmbsyfam
    \boldnotloaded\tw@
  \else \mathhexbox@{\hexnumber@\cmbsyfam}{#1}{#2}\fi}
\def\boldsymbol@#1{\let\next@\relax\let\next#1%
 \ifx\next\cdot\mathcharii@201 \else
 \ifx\next\prime{{\null\mathcharii@030 \null}}\else
 \ifx\next\lbrack\mathchar"4\bffam@5B \else
 \ifx\next\rbrack\mathchar"5\bffam@5D \else
 \ifx\next\{\mathcharii@466 \else
 \ifx\next\lbrace\mathcharii@466 \else
 \ifx\next\}\mathcharii@567 \else
 \ifx\next\rbrace\mathcharii@567 \else
 \ifx\next\surd{{\mathcharii@170}}\else
 \ifx\next\S{{\mathhexboxii@78}}\else
 \ifx\next\P{{\mathhexboxii@7B}}\else
 \ifx\next\dag{{\mathhexboxii@79}}\else
 \ifx\next\ddag{{\mathhexboxii@7A}}\else
 \DN@{\Err@{You can't use \string\boldsymbol\space with \string#1}#1}%
 \fi\fi\fi\fi\fi\fi\fi\fi\fi\fi\fi\fi\fi}
\def\boldsymbol@@#1.#2\boldsymbol@@{\classnum@#1 \count@@@\classnum@        %1
 \divide\classnum@4096 \count@\classnum@                                    %2
 \multiply\count@4096 \advance\count@@@-\count@ \count@@\count@@@           %3
 \divide\count@@@\@cclvi \count@\count@@                                    %4
 \multiply\count@@@\@cclvi \advance\count@@-\count@@@                       %5
 \divide\count@@@\@cclvi                                                    %6
 \multiply\classnum@4096 \advance\classnum@\count@@                         %7
 \ifnum\count@@@=\z@                                                        %8
  \count@"\bffam@ \multiply\count@\@cclvi
  \advance\classnum@\count@
  \DN@{\mathchar\number\classnum@}%
 \else
  \ifnum\count@@@=\@ne                                                      %9
   \ifx\undefined\cmmibfam \DN@{\boldnotloaded\@ne}%
   \else \count@\cmmibfam \multiply\count@\@cclvi
     \advance\classnum@\count@
     \DN@{\mathchar\number\classnum@}\fi
  \else
   \ifnum\count@@@=\tw@                                                    %10
     \ifx\undefined\cmbsyfam
       \DN@{\boldnotloaded\tw@}%
     \else
       \count@\cmbsyfam \multiply\count@\@cclvi
       \advance\classnum@\count@
       \DN@{\mathchar\number\classnum@}%
     \fi
  \fi
 \fi
\fi}
\newif\ifdelim@
\newcount\delimcount@
{\uccode`6=`\\ \uccode`7=`d \uccode`8=`e \uccode`9=`l
 \uppercase{\gdef\delim@#1#2#3#4#5\delim@
  {\delim@false\ifx 6#1\ifx 7#2\ifx 8#3\ifx 9#4\delim@true
   \xdef\meaning@{#5}\fi\fi\fi\fi}}}
\def\delim@@#1"#2#3#4#5#6\delim@@{\if#32%
\let\next@\relax
 \ifx\undefined\cmbsyfam \boldnotloaded\@ne
 \else \mathcharii@#2#4#5\space \fi\fi}
\def\vert{\delimiter"026A30C }
\def\Vert{\delimiter"026B30D }
\let\|\Vert

\def\boldkeydots@#1{\bold@true\let\next=#1\let\delayed@=#1\mdots@@
 \boldkey#1\bold@false}  % = required!
\def\boldsymboldots@#1{\bold@true\let\next#1\let\delayed@#1\mdots@@
 \boldsymbol#1\bold@false}
\message{Euler fonts,}

\def\frak{\mathfont@\frak}

\def\loadmathfont#1{% 
   \expandafter\font@\csname ten#1\endcsname=#110
   \expandafter\font@\csname seven#1\endcsname=#17
   \expandafter\font@\csname five#1\endcsname=#15
   \edef\next{\noexpand\alloc@@8\fam\chardef\sixt@@n
     \expandafter\noexpand\csname#1fam\endcsname}%
   \next
   \textfont\csname#1fam\endcsname \csname ten#1\endcsname
   \scriptfont\csname#1fam\endcsname \csname seven#1\endcsname
   \scriptscriptfont\csname#1fam\endcsname \csname five#1\endcsname
   \expandafter\def\csname #1\expandafter\endcsname\expandafter{%
      \expandafter\mathfont@\csname#1\endcsname}%
 \expandafter\gdef\csname load#1\endcsname{}%
}
\def\mathfont@#1{\RIfM@\expandafter\mathfont@@\expandafter#1\else
  \expandafter\nonmatherr@\expandafter#1\fi}
\def\mathfont@@#1#2{{\mathfont@@@#1{#2}}}
\def\mathfont@@@#1#2{\noaccents@
   \fam\csname\expandafter\eat@\string#1fam\endcsname
   \relax#2}
\message{math accents,}
\def\accentclass@{7}
\def\noaccents@{\def\accentclass@{0}}
\def\makeacc@#1#2{\def#1{\mathaccent"\accentclass@#2 }}
\makeacc@\hat{05E}
\makeacc@\check{014}
\makeacc@\tilde{07E}
\makeacc@\acute{013}
\makeacc@\grave{012}
\makeacc@\dot{05F}
\makeacc@\ddot{07F}
\makeacc@\breve{015}
\makeacc@\bar{016}
\def\vec{\mathaccent"017E }
\newcount\skewcharcount@
\newcount\familycount@
\def\theskewchar@{\familycount@\@ne
 \global\skewcharcount@\the\skewchar\textfont\@ne                           %1
 \ifnum\fam>\m@ne\ifnum\fam<16
  \global\familycount@\the\fam\relax
  \global\skewcharcount@\the\skewchar\textfont\the\fam\relax\fi\fi          %2
 \ifnum\skewcharcount@>\m@ne
  \ifnum\skewcharcount@<128
  \multiply\familycount@256
  \global\advance\skewcharcount@\familycount@
  \global\advance\skewcharcount@28672
  \mathchar\skewcharcount@\else
  \global\skewcharcount@\m@ne\fi\else
 \global\skewcharcount@\m@ne\fi}                                            %3
\newcount\pointcount@
\def\getpoints@#1.#2\getpoints@{\pointcount@#1 }
\newdimen\accentdimen@
\newcount\accentmu@
\def\dimentomu@{\multiply\accentdimen@ 100
 \expandafter\getpoints@\the\accentdimen@\getpoints@
 \multiply\pointcount@18
 \divide\pointcount@\@m
 \global\accentmu@\pointcount@}
\def\Makeacc@#1#2{\def#1{\RIfM@\DN@{\mathaccent@
 {"\accentclass@#2 }}\else\DN@{\nonmatherr@{#1}}\fi\next@}}
\def\unbracefonts@{\let\Cal@\Cal@@\let\roman@\roman@@\let\bold@\bold@@
 \let\slanted@\slanted@@}
\def\mathaccent@#1#2{\ifnum\fam=\m@ne\xdef\thefam@{1}\else
 \xdef\thefam@{\the\fam}\fi                                                 %1
 \accentdimen@\z@                                                           %2
 \setboxz@h{\unbracefonts@$\m@th\fam\thefam@\relax#2$}%                     %3
 \ifdim\accentdimen@=\z@\DN@{\mathaccent#1{#2}}%                            %4
  \setbox@ne\hbox{\unbracefonts@$\m@th\fam\thefam@\relax#2\theskewchar@$}% %5a
  \setbox\tw@\hbox{$\m@th\ifnum\skewcharcount@=\m@ne\else
   \mathchar\skewcharcount@\fi$}%                                          %5b
  \global\accentdimen@\wd@ne\global\advance\accentdimen@-\wdz@
  \global\advance\accentdimen@-\wd\tw@                                     %5c
  \global\multiply\accentdimen@\tw@
  \dimentomu@\global\advance\accentmu@\@ne                                 %5d
 \else\DN@{{\mathaccent#1{#2\mkern\accentmu@ mu}%
    \mkern-\accentmu@ mu}{}}\fi                                             %6
 \next@}\Makeacc@\Hat{05E}
\Makeacc@\Check{014}
\Makeacc@\Tilde{07E}
\Makeacc@\Acute{013}
\Makeacc@\Grave{012}
\Makeacc@\Dot{05F}
\Makeacc@\Ddot{07F}
\Makeacc@\Breve{015}
\Makeacc@\Bar{016}
\def\Vec{\RIfM@\DN@{\mathaccent@{"017E }}\else
 \DN@{\nonmatherr@\Vec}\fi\next@}
\def\accentedsymbol#1#2{\csname newbox\expandafter\endcsname
  \csname\expandafter\eat@\string#1@box\endcsname
 \expandafter\setbox\csname\expandafter\eat@
  \string#1@box\endcsname\hbox{$\m@th#2$}\define
  #1{\copy\csname\expandafter\eat@\string#1@box\endcsname{}}}
\message{roots,}
\def\sqrt#1{\radical"270370 {#1}}
\let\underline@\underline
\let\overline@\overline
\def\underline#1{\underline@{#1}}
\def\overline#1{\overline@{#1}}
\Invalid@\leftroot
\Invalid@\uproot
\newcount\uproot@
\newcount\leftroot@
\def\root{\relaxnext@
  \DN@{\ifx\next\uproot\let\next@\nextii@\else
   \ifx\next\leftroot\let\next@\nextiii@\else
   \let\next@\plainroot@\fi\fi\next@}%
  \DNii@\uproot##1{\uproot@##1\relax\FN@\nextiv@}%
  \def\nextiv@{\ifx\next\space@\DN@. {\FN@\nextv@}\else
   \DN@.{\FN@\nextv@}\fi\next@.}%
  \def\nextv@{\ifx\next\leftroot\let\next@\nextvi@\else
   \let\next@\plainroot@\fi\next@}%
  \def\nextvi@\leftroot##1{\leftroot@##1\relax\plainroot@}%
   \def\nextiii@\leftroot##1{\leftroot@##1\relax\FN@\nextvii@}%
  \def\nextvii@{\ifx\next\space@
   \DN@. {\FN@\nextviii@}\else
   \DN@.{\FN@\nextviii@}\fi\next@.}%
  \def\nextviii@{\ifx\next\uproot\let\next@\nextix@\else
   \let\next@\plainroot@\fi\next@}%
  \def\nextix@\uproot##1{\uproot@##1\relax\plainroot@}%
  \bgroup\uproot@\z@\leftroot@\z@\FN@\next@}
\def\plainroot@#1\of#2{\setbox\rootbox\hbox{$\m@th\scriptscriptstyle{#1}$}%
 \mathchoice{\r@@t\displaystyle{#2}}{\r@@t\textstyle{#2}}
 {\r@@t\scriptstyle{#2}}{\r@@t\scriptscriptstyle{#2}}\egroup}
\def\r@@t#1#2{\setboxz@h{$\m@th#1\sqrt{#2}$}%
 \dimen@\ht\z@\advance\dimen@-\dp\z@
 \setbox@ne\hbox{$\m@th#1\mskip\uproot@ mu$}\advance\dimen@ 1.667\wd@ne
 \mkern-\leftroot@ mu\mkern5mu\raise.6\dimen@\copy\rootbox
 \mkern-10mu\mkern\leftroot@ mu\boxz@}
\def\boxed#1{\setboxz@h{$\m@th\displaystyle{#1}$}\dimen@.4\ex@
 \advance\dimen@3\ex@\advance\dimen@\dp\z@
 \hbox{\lower\dimen@\hbox{%
 \vbox{\hrule height.4\ex@
 \hbox{\vrule width.4\ex@\hskip3\ex@\vbox{\vskip3\ex@\boxz@\vskip3\ex@}%
 \hskip3\ex@\vrule width.4\ex@}\hrule height.4\ex@}%
 }}}
\message{commutative diagrams,}
\let\ampersand@\relax
\newdimen\minaw@
\minaw@11.11128\ex@
\newdimen\minCDaw@
\minCDaw@2.5pc
\def\minCDarrowwidth#1{\RIfMIfI@\onlydmatherr@\minCDarrowwidth
 \else\minCDaw@#1\relax\fi\else\onlydmatherr@\minCDarrowwidth\fi}
\newif\ifCD@
\def\CD{\bgroup\vspace@\relax\iffalse{\fi\let\ampersand@&\iffalse}\fi
 \CD@true\vcenter\bgroup\Let@\tabskip\z@skip\baselineskip20\ex@
 \lineskip3\ex@\lineskiplimit3\ex@\halign\bgroup
 &\hfill$\m@th##$\hfill\crcr}
\def\endCD{\crcr\egroup\egroup\egroup}
\newdimen\bigaw@
\atdef@>#1>#2>{\ampersand@                                                  %1
 \setboxz@h{$\m@th\ssize\;{#1}\;\;$}%                                       %2
 \setbox@ne\hbox{$\m@th\ssize\;{#2}\;\;$}%                                  %3
 \setbox\tw@\hbox{$\m@th#2$}%                                               %4
 \ifCD@\global\bigaw@\minCDaw@\else\global\bigaw@\minaw@\fi                 %5
 \ifdim\wdz@>\bigaw@\global\bigaw@\wdz@\fi
 \ifdim\wd@ne>\bigaw@\global\bigaw@\wd@ne\fi                                %6
 \ifCD@\enskip\fi                                                           %7
 \ifdim\wd\tw@>\z@
  \mathrel{\mathop{\hbox to\bigaw@{\rightarrowfill@\displaystyle}}%
    \limits^{#1}_{#2}}%                                                     %8
 \else\mathrel{\mathop{\hbox to\bigaw@{\rightarrowfill@\displaystyle}}%
    \limits^{#1}}\fi                                                        %9
 \ifCD@\enskip\fi                                                          %10
 \ampersand@}                                                              %11
\atdef@<#1<#2<{\ampersand@\setboxz@h{$\m@th\ssize\;\;{#1}\;$}%
 \setbox@ne\hbox{$\m@th\ssize\;\;{#2}\;$}\setbox\tw@\hbox{$\m@th#2$}%
 \ifCD@\global\bigaw@\minCDaw@\else\global\bigaw@\minaw@\fi
 \ifdim\wdz@>\bigaw@\global\bigaw@\wdz@\fi
 \ifdim\wd@ne>\bigaw@\global\bigaw@\wd@ne\fi
 \ifCD@\enskip\fi
 \ifdim\wd\tw@>\z@
  \mathrel{\mathop{\hbox to\bigaw@{\leftarrowfill@\displaystyle}}%
       \limits^{#1}_{#2}}\else
  \mathrel{\mathop{\hbox to\bigaw@{\leftarrowfill@\displaystyle}}%
       \limits^{#1}}\fi
 \ifCD@\enskip\fi\ampersand@}
\begingroup
 \catcode`\~=\active \lccode`\~=`\@
 \lowercase{%
  \global\atdef@)#1)#2){~>#1>#2>}
  \global\atdef@(#1(#2({~<#1<#2<}}
\endgroup
\atdef@ A#1A#2A{\llap{$\m@th\vcenter{\hbox
 {$\ssize#1$}}$}\Big\uparrow\rlap{$\m@th\vcenter{\hbox{$\ssize#2$}}$}&&}
\atdef@ V#1V#2V{\llap{$\m@th\vcenter{\hbox
 {$\ssize#1$}}$}\Big\downarrow\rlap{$\m@th\vcenter{\hbox{$\ssize#2$}}$}&&}
\atdef@={&\enskip\mathrel
 {\vbox{\hrule width\minCDaw@\vskip3\ex@\hrule width
 \minCDaw@}}\enskip&}
\atdef@|{\Big\Vert&&}
\atdef@\vert{\Big\Vert&&}
\def\pretend#1\haswidth#2{\setboxz@h{$\m@th\scriptstyle{#2}$}\hbox
 to\wdz@{\hfill$\m@th\scriptstyle{#1}$\hfill}}
\message{poor man's bold,}
\def\pmb{\RIfM@\expandafter\mathpalette\expandafter\pmb@\else
 \expandafter\pmb@@\fi}
\def\pmb@@#1{\leavevmode\setboxz@h{#1}%
   \dimen@-\wdz@
   \kern-.5\ex@\copy\z@
   \kern\dimen@\kern.25\ex@\raise.4\ex@\copy\z@
   \kern\dimen@\kern.25\ex@\box\z@
}
\def\binrel@@#1{\ifdim\wd2<\z@\mathbin{#1}\else\ifdim\wd\tw@>\z@
 \mathrel{#1}\else{#1}\fi\fi}
\newdimen\pmbraise@
%      Note: because of the use of \mathpalette, if \pmb@ is
%      applied to a single math italic character (or a single
%      character from some other slanted math font), the italic
%      correction will be added.  This will cause subscripts
%      to fall too far away from the character in some
%      cases, e.g., $\pmb{T}_1$ or $\pmb{\Cal T}_1$.
\def\pmb@#1#2{\setbox\thr@@\hbox{$\m@th#1{#2}$}%
 \setbox4\hbox{$\m@th#1\mkern.5mu$}\pmbraise@\wd4\relax
 \binrel@{#2}%
 \dimen@-\wd\thr@@
   \binrel@@{%
   \mkern-.8mu\copy\thr@@
   \kern\dimen@\mkern.4mu\raise\pmbraise@\copy\thr@@
   \kern\dimen@\mkern.4mu\box\thr@@
}}
\def\documentstyle#1{\W@{}\input #1.sty\relax}
\message{syntax check,}
\font\dummyft@=dummy
\fontdimen1 \dummyft@=\z@
\fontdimen2 \dummyft@=\z@
\fontdimen3 \dummyft@=\z@
\fontdimen4 \dummyft@=\z@
\fontdimen5 \dummyft@=\z@
\fontdimen6 \dummyft@=\z@
\fontdimen7 \dummyft@=\z@
\fontdimen8 \dummyft@=\z@
\fontdimen9 \dummyft@=\z@
\fontdimen10 \dummyft@=\z@
\fontdimen11 \dummyft@=\z@
\fontdimen12 \dummyft@=\z@
\fontdimen13 \dummyft@=\z@
\fontdimen14 \dummyft@=\z@
\fontdimen15 \dummyft@=\z@
\fontdimen16 \dummyft@=\z@
\fontdimen17 \dummyft@=\z@
\fontdimen18 \dummyft@=\z@
\fontdimen19 \dummyft@=\z@
\fontdimen20 \dummyft@=\z@
\fontdimen21 \dummyft@=\z@
\fontdimen22 \dummyft@=\z@
\def\fontlist@{\\{\tenrm}\\{\sevenrm}\\{\fiverm}\\{\teni}\\{\seveni}%
 \\{\fivei}\\{\tensy}\\{\sevensy}\\{\fivesy}\\{\tenex}\\{\tenbf}\\{\sevenbf}%
 \\{\fivebf}\\{\tensl}\\{\tenit}}
\def\font@#1=#2 {\rightappend@#1\to\fontlist@\font#1=#2 }
\def\dodummy@{{\def\\##1{\global\let##1\dummyft@}\fontlist@}}
\def\nopages@{\output{\setbox\z@\box\@cclv \deadcycles\z@}%
 \alloc@5\toks\toksdef\@cclvi\output}
\let\galleys\nopages@
\newif\ifsyntax@
\newcount\countxviii@
\def\syntax{\syntax@true\dodummy@\countxviii@\count18
 \loop\ifnum\countxviii@>\m@ne\textfont\countxviii@=\dummyft@
 \scriptfont\countxviii@=\dummyft@\scriptscriptfont\countxviii@=\dummyft@
 \advance\countxviii@\m@ne\repeat                                           %1
 \dummyft@\tracinglostchars\z@\nopages@\frenchspacing\hbadness\@M}
\def\first@#1#2\end{#1}
\def\printoptions{\W@{Do you want S(yntax check),
  G(alleys) or P(ages)?}%
 \message{Type S, G or P, followed by <return>: }%
 \begingroup % to localize the following change to \endlinechar:
 \endlinechar\m@ne % to prevent a space or \par in \ans@ from ^^M
 \read\m@ne to\ans@
%  Define \ans@ to uppercase itself, and default to P if the user
%  just pressed <return>.
 \edef\ans@{\uppercase{\def\noexpand\ans@{%
   \expandafter\first@\ans@ P\end}}}%
%  Cast the new definition of \ans@ outside the group
 \expandafter\endgroup\ans@
 \if\ans@ P% fine, no action needs to be taken
 \else \if\ans@ S\syntax
 \else \if\ans@ G\galleys
 \else\message{? Unknown option: \ans@; using the `pages' option.}%
 \fi\fi\fi}
\def\alloc@#1#2#3#4#5{\global\advance\count1#1by\@ne
 \ch@ck#1#4#2\allocationnumber=\count1#1
 \global#3#5=\allocationnumber
 \ifalloc@\wlog{\string#5=\string#2\the\allocationnumber}\fi}
\def\document{\def\alloclist@{}\def\fontlist@{}}

\let\footnote\undefined
\let\=\undefined
\let\>\undefined

\catcode`\@=\active
\message{... finished}
%\endinput

\nopagenumbers
%This command suppresses the printing of page numbers.
%You should number the pages with blue pencil in upper right corner. 
%
%THE FOLLOWING THREE COMMANDS LEAVE SOME SPACE AT THE TOP OF THE LEAD PAGE. 
%(the command "\vskip 4 truecm" actually results in about 4.5 cm of empty
%space at the top, or about 19.6%).  The publisher will probably reset the 
%chapter heading (your title and by-line), but you should follow my 
%19-20% prescription anyway!  In my design I am following the Les Houches 
%lecture notes volume produced by Nova.   If you have LOTS of authors and 
%by-lines you may want to allow a bit less space at the top (e.g. if you 
%have 3 or more sets of authors and institutions).
\topinsert
\vskip 3.2 truecm
\endinsert
\centerline{\bigbfont HIERARCHICAL MEAN-FIELD THEORIES}
%If your title is only one line long, put a % before the 2nd title line.
%If your title is longer than two lines, continue thus:
%\vskip 6 truept
%\centerline{\bigbfont AND THE TITLE GOES ON AND ON}
%Don't forget to remove the % sign from the preceding line if you use it! 
\vskip 20 truept
%Now comes your by-line with institutional addresses.  
\centerline{\bigifont G. Ortiz and C.D. Batista}
\vskip 8 truept
\centerline{\bigrfont Theoretical Division, 
Los Alamos National Laboratory} 
\vskip 2 truept
\centerline{\bigrfont Los Alamos, NM 87545, USA} 
%In case of multiple institutions, use the following lines, iterated
%as necessary. 
%\vskip 14 truept
%\centerline{\bigifont AuthorX, AuthorY, and AuthorZ}
%\vskip 8 truept
%\centerline{\bigrfont Department of Physics, Generic University}
%\vskip 2 truept
%\centerline{\bigrfont Generic Address Line 2} 
\vskip 1.8 truecm

\centerline{\bf 1.  INTRODUCTION TO COMPLEX QUANTUM ORDERS}
\vskip 12 truept

One of the most challenging and interesting phenomena in modern
condensed matter physics is the one emerging from strongly coupled
systems as a result of competing interactions. Indeed, the multiplicity
of distinct and novel quantum phases observed experimentally confront
us with new paradigms that question our understanding of the
fundamental principles behind such complex phenomena. For example, 
whether the mechanism controlling the coexistence and/or competition
between magnetism and superconductivity or Bose-Einstein condensation
has the same physical origin in different classes of materials is still
an open question [1]. It is believed that the physics in these
materials is strongly influenced by their proximity to quantum phase
transitions and, in particular, to quantum criticality. A quantum phase
transition is characterized by the qualitative changes of the
macroscopic state of the system induced by tuning parameters of its
Hamiltonian. On general grounds, the very notion and
nature of entanglement is at the core of the problem [2]. 

From the theoretical viewpoint the hurdle is in the presence of
non-linear couplings, non-perturbative phenomena, and a panoply of
competing quantum orders. These systems happened to be strongly
correlated since no obvious small coupling constant exists, and
consequently exhibit high sensitivity to small parameter changes. It is
then clear the importance of developing a methodology, based on
qualitatively new concepts, that treats all possible competing orders
on an equal footing with no privileged fixed-point phenomenon. Despite
great advances there is a lack of a systematic and reliable methodology
to study and predict the behavior of these complex systems. It is a
purpose of this work to present a promising step in that direction. 

In the quantum description of matter, a physical system is naturally
associated with a {\it language} of operators. We have previously 
developed an algebraic framework for interacting quantum systems that
let us study complex phenomena characterized by the coexistence and
competition of various broken symmetry states [3-5], and proved a
theorem that allowed us to connect all possible languages used in the
quantum description of matter. Connecting the various languages through
isomorphic mappings enable us to relate seemingly different physical
phenomena, unveil hidden symmetries (i.e., uncover the {\it accidental
degeneracies} of the original physical system), and, in some limiting
cases, obtain the exact spectrum of the problem (or of a set of
orthogonal subspaces). The ultimate goal was to use that framework to
explore those unconventional complex states of matter from a unified
perspective. Given the space limitations we cannot review these
concepts but refer the reader to a recent review article on the subject
[5]. 

The modern theory of phase transitions starts with Landau's pioneering
work in 1937 [6]. One of his achievements was the realization
of the fundamental relation between spontaneous symmetry breaking and
the order parameter (OP) that measures this violation, thus giving
simple prescriptions to describe order in terms of irreducible
representations of the symmetry group involved. Another was the
development of a phenomenological calculational scheme to study the
%behavior of systems near a phase transition. Over the years Landau's
behavior of systems near a phase transition. Landau's
theory has been successfully applied to study phase transitions where
thermal fluctuations are most relevant. Certainly, the theory was not
designed to study zero-temperature (quantum) phase transitions. 

In previous work [3,4] we outlined a framework to identify OPs based
upon isomorphic mappings to a {\it hierarchical language} (HL) defined
by the set of operators which in the fundamental representation (of
dimension $D$) has the largest number of symmetry generators of the
group. {\it Any} local operator can be expressed as a {\it linear}
combination of the generators of the HL. The building of the HL depends
upon the dimension $D$ of the local Hilbert space, ${\Cal H}_{\bj}$,
modeling the physical phenomena. For instance, if one is modeling a
doped antiferromagnetic (AF) insulator with a $t$-$J$ Hamiltonian [7],
then $D$=3 (i.e., there are three possible states per site) and a HL is
generated by a basis of the Lie algebra $su(3)$ in the fundamental
representation [3,4]. As explained and proved in Refs. [3-5], there is
always a HL associated to each physical problem. These ideas complement
Landau's concept of an OP  providing a mechanism to reveal them,
something that is outside the groundwork of his theory. Indeed,
Landau's theory does not say what the OPs should be in a general
situation.

As mentioned above, these isomorphic mappings not only unveil hidden
symmetries of the original physical system but also manifestly
establish equivalences between seemingly unrelated physical phenomena.
Nonetheless, this is not sufficient to determine the {\it exact} phase
diagram of the problem: One has to resort to either numerical
simulations with their well-known limitations or, as will be shown in
the present paper, to a {\it guided} approximation which at least
preserves the qualitative nature of the possible thermodynamic states.
A key observation in this regard is the fact that typical model
Hamiltonian operators written in the HL become quadratic in the
symmetry generators of the hierarchical group, and this result is
independent of the group of symmetries of the Hamiltonian. 

This latter result suggests a simple approximation, based upon group
theoretical grounds, which deals with competing orders on an equal
footing and will be termed {\it hierarchical mean-field theory} (HMFT).
%In a sense, that will become clear below, HMFT constitutes the {\it
%optimum} mean-field (MF) or saddle-point solution that approximates the
%energy and correlation functions of the original problem. 
The HMFT is distinctly suitable when the various phases displayed by a
system are the result of competing interactions and non-linear
couplings of their constituents matter fields. 

%The main questions we address in the present paper are:
%
%--- What can be done when there is no exact (or quasi-exact) solution? 
%
%--- Is there a concept of optimum mean-field approximation?
%
%--- Can we build reliable phase diagrams?
%
%The answers to these questions will be connected to the term HMFT.

\vskip 1.8 truecm

\centerline{\bf 2.  HIERARCHICAL MEAN-FIELD THEORY AT WORK}
\vskip 12 truept

In the rest of the paper we will expand on a technique to build
approximate phase diagrams, dubbed HMFT. To avoid excess of tedious
formalism we will describe the methodology by example. We have chosen
two representative examples. The first displays two quantum phases
separated by a bosonic {\it metal-insulator} transition induced by
particle interactions. The second shows an AF, a superfluid and, in 
addition, a coexistence phase as a function of the particle density and
temperature. 

\vskip 12 truept
\centerline{\bf 2.1  A Superfluid-Insulator Transition}
\vskip 12 truept

Let us determine the zero temperature phase diagram of a very simple
model displaying a Mott insulating to superfluid transition. The model
we refer to is a  modification of the Bose-Hubbard Hamiltonian [8]
($U,t>0$)
$$
\eqalignno{
H&=-2t \! \sum_{\langle \text{\bf i},\text{\bf j} \rangle} (
{{g}}^\dagger_{\text{\bf i}} {{g}}^{\;}_{\text{\bf j}} +
{{g}}^\dagger_{\text{\bf j}} {{g}}^{\;}_{\text{\bf i}} ) + U
\sum_{\text{\bf j}}  ({n}_{\text{\bf j}} - \bar{n}+1)({n}_{\text{\bf
j}} - \bar{n})   + \sum_{\text{\bf j}} \epsilon_{\text{\bf j}} \
{{n}}_{\text{\bf j}} \ , \cr
&=-2t \! \sum_{\langle \text{\bf i},\text{\bf j} \rangle} (
{{g}}^\dagger_{\text{\bf i}} {{g}}^{\;}_{\text{\bf j}} +
{{g}}^\dagger_{\text{\bf j}} {{g}}^{\;}_{\text{\bf i}} ) + U
\sum_{\text{\bf j}}  {n}_{\text{\bf j}} ({n}_{\text{\bf j}} - 1)   +
\sum_{\text{\bf j}} \mu_{\bj} \ {{n}}_{\text{\bf j}} + N_s U
\bar{n}(\bar{n}-1)\ , & (1)
}
$$
where $\langle \text{\bf i},\text{\bf j} \rangle$ stands for
nearest-neighbor sites (bond) in an otherwise regular $N_s$-sites
lattice of coordination $\text{\sl z}$, $\mu_{\bj} =
2U(1-\bar{n})+\epsilon_{\text{\bf j}}$, and the $g$-particles, instead
of  canonical bosons, represent bosons satisfying the following
commutation relations ($(g^\dagger_{\bj})^\dagger=g^{\;}_{\bj},
{n}_{\bj}  \neq g^\dagger_{\bj}g^{\;}_{\bj}$) 
$$
\cases{
{[} {g}^{\;}_{\bi}, {g}^{\;}_{\bj} {]} =
{[} {g}^\dagger_{\bi}, {g}^\dagger_{\bj} {]} = 0  \ , \cr
{[} {g}^{\;}_{\bi}, {g}^\dagger_{\bj} {]} = \delta_{\bi\bj} \  
\bar{n}(\bar{n}-{n}_{\bj}) \ , \ \ \ 
{[}n_{\bi}, {g}^\dagger_{\bj} {]}=\delta_{\bi\bj} \ {g}^\dagger_{\bj}  
\ , 
} \eqno(2)
$$
defining an algebra, with $\bar{n}\geq 1$ a positive integer and the
nilpotency condition $(g^\dagger_{\bj})^{p+1}=(g^\dagger_{\bj})^3=0$,
meaning that one can accommodate up to $p=2$ particles per mode $\bj$.
This indicates that the local Hilbert space ${\Cal
H}_{\bj}$ is three-dimensional. A possible basis is $\{
|\bar{n}-1\rangle, |\bar{n}\rangle, |\bar{n}+1\rangle\}$, and the
operators acting on this basis
$$
\matrix{
g^{\;}_{\bj}|\bar{n}-1\rangle= 0 \ , \hfill & g^{\dagger}_{\bj}
|\bar{n}-1\rangle  = \sqrt{\bar{n}} \ |\bar{n}\rangle \ , \hfill &
n^{\;}_{\bj}|\bar{n}-1\rangle =  (\bar{n}-1)|\bar{n}-1\rangle , \hfill
\cr
g^{\;}_{\bj}|\bar{n}\rangle = \sqrt{\bar{n}} \ |\bar{n}-1\rangle \ ,
\hfill  & g^{\dagger}_{\bj}|\bar{n}\rangle  = \sqrt{\bar{n}} \
|\bar{n}+1\rangle \ , \hfill & n^{\;}_{\bj}|\bar{n}\rangle = 
\bar{n}|\bar{n}\rangle , \hfill \cr
g^{\;}_{\bj}|\bar{n}+1\rangle = \sqrt{\bar{n}} \ |\bar{n}\rangle \ ,
\hfill  & g^{\dagger}_{\bj}|\bar{n}+1\rangle = 0 \ ,\hfill  &
n^{\;}_{\bj}|\bar{n}+1\rangle =  (\bar{n}+1)|\bar{n}+1\rangle ,\hfill 
\cr}\eqno(3)
$$
behave as harmonic-oscillator-like operators acting on a
three-dimensional space. Since $D=3$ one can easily determine the 
isomorphic mapping between the $g$-particle and the $su(2)$ algebra in
the $S=1$ representation. The transformation is given by
$$
\cases{
\sqrt{\frac{2}{\bar{n}}} \ {g}^\dagger_{\bj} = S^+_{\bj}  \ , \cr
\sqrt{\frac{2}{\bar{n}}} \ {g}^{\;}_{\bj} = S^-_{\bj}  \ , \cr
n_{\bj}-\bar{n}= S^z_{\bj} \ ,
} \eqno(5)
$$
while the Hamiltonian operator in the spin language reads
$$
H=-t \bar{n} \! \sum_{\langle {\bi},{\bj} \rangle} \left (
S^+_{\bi} S^-_{\bj} + S^-_{\bi} S^+_{\bj}\right ) + U \sum_{\bj} 
(S^z_{\bj})^2 + \sum_{{\bj}} (U+\epsilon_{\bj}) \ S^z_{\bj} \ .\eqno(6)
$$
Before translating the problem into the most fundamental language,
i.e., the HL, let us make a detour and re-express $H$ using the
spin-1/2 ($\sigma= \uparrow,\downarrow$) Jordan-Wigner bosons [9],
$$
\cases{
S^+_{\bj}= \sqrt{2} \ (\bar{b}^{\dagger}_{{\bj} \uparrow} + 
\bar{b}^{\;}_{{\bj}\downarrow}) \ , \cr
S^-_{\bj}= \sqrt{2} \ (\bar{b}^{\;}_{{\bj} \uparrow} + 
\bar{b}^{\dagger}_{{\bj}\downarrow})  \ , \cr
S^z_{\bj}= \bar{n}_{{\bj} \uparrow}-\bar{n}_{{\bj} \downarrow}\ ,
} \eqno(7)
$$
$$
H=-2t \bar{n} \!\! \sum_{\langle {\bi},{\bj} \rangle,\sigma} (
{\bar{b}}^\dagger_{{\bi} \sigma} {\bar{b}}^{\;}_{{\bj} \sigma} +
{\bar{b}}^\dagger_{{\bi} \sigma} {\bar{b}}^{\dagger}_{{\bj}
\bar{\sigma}} + \text{\rm H.c.} ) + U \sum_{\bj} \bar{n}_{\bj} + 
\sum_{\bj} (U+\epsilon_{\bj}) (\bar{n}_{\bj \uparrow} - \bar{n}_{\bj
\downarrow}) \ ,\eqno(8)
$$
where the number operator ${\bar {n}}_{\text{\bf j}}=\bar{n}_{\text{\bf
j} \uparrow} + \bar{n}_{\text{\bf j} \downarrow}$ ($\bar{n}_{\text{\bf
j} \sigma}={\bar {b}}^\dagger_{\text{\bf j} \sigma} {\bar
{b}}^{\;}_{\text{\bf j} \sigma}$), and  the algebra satisfied by these
hard-core bosons is [5]: $[\bar{b}^{\;}_{\text{\bf
i}\sigma},\bar{b}^{\;}_{\text{\bf j}\sigma'} ]=0$,
$[\bar{b}^{\;}_{\text{\bf i}\sigma},\bar{b}^{\dagger}_{\text{\bf
j}\sigma'} ]= \delta_{\text{\bf ij}} (1-2\bar{n}_{\text{\bf j}\sigma}-
\bar{n}_{\text{\bf j}\bar{\sigma}})$ (if $\sigma=\sigma'$), or 
$-\delta_{\text{\bf ij}} \bar{b}^{\dagger}_{\text{\bf j}\sigma'} 
\bar{b}^{\;}_{\text{\bf j}\sigma}$ (if $\sigma \neq \sigma'$). 

As explained in the introduction, the first step in determining its
phase diagram consists of re-writing $H$ in a HL. The latter is
realized by $SU(3)$-spin generators in the fundamental representation,
and its mapping to the hard-core boson language can be compactly
written as [3]
$$
{\Cal S}(\text{\bf j})= \pmatrix{ \frac{2}{3} - \bar{n}_{\text{\bf j}}
&\bar{b}^{\;}_{\text{\bf j} \uparrow} & \bar{b}^{\;}_{\text{\bf j}
\downarrow} \cr &&\cr \bar{b}^{\dagger}_{\text{\bf j}
\uparrow}&\bar{n}_{\text{\bf j} \uparrow} -\frac{1}{3}&
\bar{b}^\dagger_{\text{\bf j} \uparrow} \bar{b}^{\;}_{\text{\bf j}
\downarrow} \cr &&\cr \bar{b}^{\dagger}_{\text{\bf j}
\downarrow}&\bar{b}^\dagger_{\text{\bf j} \downarrow}
\bar{b}^{\;}_{\text{\bf j} \uparrow}&\bar{n}_{\text{\bf j}
\downarrow}-\frac{1}{3} } , \
\tilde{\Cal S}({\bj})= \pmatrix{ \frac{2}{3} - \bar{n}_{{\bj}} &
-\bar{b}^{\dagger}_{{\bj} \downarrow}& -\bar{b}^{\dagger}_{{\bj}
\uparrow} \cr &&\cr -\bar{b}^{\;}_{{\bj} \downarrow}&\bar{n}_{{\bj}
\downarrow} -\frac{1}{3}& \bar{b}^\dagger_{{\bj} \uparrow}
\bar{b}^{\;}_{{\bj} \downarrow} \cr &&\cr -\bar{b}^{\;}_{{\bj}
\uparrow}&\bar{b}^\dagger_{{\bj}  \downarrow} \bar{b}^{\;}_{{\bj}
\uparrow}&\bar{n}_{{\bj} \uparrow}-\frac{1}{3} } .\eqno(9)
%\label{spinsu3}
$$
The three components $s^{z}_{\text{\bf j}}=(\bar{n}_{\text{\bf j}\uparrow}
-\bar{n}_{\text{\bf j}\downarrow})/2$, $s^{+}_{\text{\bf
j}}=\bar{b}^\dagger_{\text{\bf j} \uparrow} \bar{b}^{\;}_{\text{\bf j}
\downarrow}$ and $s^{-}_{\text{\bf j}}=\bar{b}^\dagger_{\text{\bf j}
\downarrow} \bar{b}^{\;}_{\text{\bf j} \uparrow}$  generate the spin
$su(2)$ subalgebra, i.e., they are the components of the local
magnetization. The five additional components correspond to the
Bose-Einstein condensate and the charge density wave local OPs. In the
HL, $H$ represents a Heisenberg-like Hamiltonian [10] in the presence of
an external magnetic field ($J_{\mu \nu}=J_{\nu \mu}$)
$$
H \!= \!\!\sum_{\langle {\bi},{\bj} \rangle} J_{\mu \nu} ({\Cal S}^{\mu
\nu}({\bi}) {\Cal S}^{\nu \mu}({\bj}) - {\Cal S}^{\mu
\nu}({\bi}) \tilde{\Cal S}^{\nu \mu}({\bj}) ) - U \sum_{{\bj}}
{\Cal S}^{00}({\bj})  +  \sum_{\bj} (U+\epsilon_{\bj}) ({\Cal
S}^{11}({\bj})-{\Cal S}^{22}({\bj})) ,\eqno(10)
$$
with $J_{00}=J_{11}=J_{22}=J_{12}=0$, $J_{01}=J_{02}=-2t\bar{n}$. Note
that through the  mapping we transformed an interacting problem into
another problem that is quadratic in the basis of the algebra $su(3)$
in the fundamental representation, but which is not necessarily $SU(3)$
symmetric.  The idea behind the HMFT is to perform an approximation
which deals with all possible local OPs on an equal footing with no
privileged {\it symmetry axes} and, hopefully, retains the qualitative
topology of the phase diagram. With $H$ written in the HL one
immediately  realizes that the simplest HMFT can be achieved if we
re-write $H$ in terms of $SU(3)$ Schwinger-Wigner (SW) bosons (3
flavors $\alpha=\downarrow,0,\uparrow$) [7]. The mapping is expressed
as
$$
{\Cal S}({\bj})= \pmatrix{ {n}_{{\bj}0}-\frac{1}{3}&
{b}^\dagger_{{\bj}0} {b}^{\;}_{{\bj} \uparrow}&
{b}^\dagger_{{\bj}0} {b}^{\;}_{{\bj} \downarrow}\cr &&\cr
{b}^\dagger_{{\bj}\uparrow}{b}^{\;}_{{\bj}0}& 
{n}_{{\bj} \uparrow} -\frac{1}{3}&
{b}^\dagger_{{\bj} \uparrow} {b}^{\;}_{{\bj} \downarrow} \cr &&\cr
{b}^\dagger_{{\bj}\downarrow}{b}^{\;}_{{\bj}0}&
{b}^\dagger_{{\bj} \downarrow}{b}^{\;}_{{\bj} \uparrow}&
{n}_{{\bj} \downarrow}-\frac{1}{3}
} , \
\tilde{\Cal S}({\bj})= \pmatrix{ {n}_{{\bj}0}-\frac{1}{3}&
-{b}^\dagger_{{\bj}\downarrow} {b}^{\;}_{{\bj} 0}&
-{b}^\dagger_{{\bj}\uparrow} {b}^{\;}_{{\bj}0}\cr &&\cr
-{b}^\dagger_{{\bj}0}{b}^{\;}_{{\bj}\downarrow}& 
{n}_{{\bj} \downarrow} -\frac{1}{3}&
{b}^\dagger_{{\bj} \uparrow} {b}^{\;}_{{\bj} \downarrow} \cr &&\cr
-{b}^\dagger_{{\bj}0}{b}^{\;}_{{\bj}\uparrow}&
{b}^\dagger_{{\bj} \downarrow}{b}^{\;}_{{\bj} \uparrow}&
{n}_{{\bj} \uparrow}-\frac{1}{3} 
} ,\eqno(11)
$$
with the SW bosons ${b}^\dagger_{\text{\bf j}\alpha}$ satisfying the
constraint ${n}_{\text{\bf j} \downarrow}+{n}_{\text{\bf j}0}
+{n}_{\text{\bf j} \uparrow}=1$. In this way
$$
H=\! - 2t\bar{n} \! \sum_{\langle {\bi},{\bj} \rangle, \sigma}  (
\text{\bf :} A^\dagger_{\sigma{\bi\bj}}A^{\;}_{\sigma{\bi\bj}}\text{\bf
:} -  B^\dagger_{\sigma{\bi\bj}}B^{\;}_{\sigma{\bi\bj}} ) - U 
\sum_{{\bj}}n_{{\bj}0} +   \sum_{\bj} (U+\epsilon_{\bj}) ({n}_{\bj
\uparrow} - {n}_{\bj \downarrow}) \ ,\eqno(12)
$$
\vskip -4 truept

\hskip -22 truept 
where $\text{\bf : :}$ denotes normal ordering and
$$
\cases{
A^\dagger_{\sigma{\bi\bj}}= {b}^\dagger_{{\bi} \sigma} {
b}^{\;}_{{\bj} 0} + {b}^\dagger_{{\bi} 0} {
b}^{\;}_{{\bj} \bar{\sigma}}, \cr
B^\dagger_{\sigma{\bi\bj}}= {b}^\dagger_{{\bi} \sigma} {
b}^\dagger_{{\bj}0} - {b}^\dagger_{{\bi}0} {
b}^\dagger_{{\bj} \sigma} .
} \eqno(13)
$$
If $\epsilon_{\bj}$ is translationally invariant, then 
$S^z=\sum_{\bj} S^z_{\bj}=\sum_{\bj}({n}_{\bj \uparrow} - {n}_{\bj
\downarrow})$ is a constant of motion ($[H,S^z]=0$). In the following
we  will only consider the case $S^z=0$.

Since the $su(N)$ languages provide a complete set of HLs [11], any
model Hamiltonian can be written in a similar fashion once we identify
the appropriate HL and apply the corresponding SW mapping in the {\it
fundamental representation} (the ordering operators will, of course,
have a different meaning and algebraic expressions). The key point is
that the Hamiltonian operator in the HL becomes quadratic in the
symmetry generators of the hierarchical group ($SU(3)$ in the present
case). 

The idea behind any MF approximation is to disentangle interaction
terms into quadratic ones replacing some of the elementary mode
operators by their mean value.
%in such a way
%that the resulting MF Hamiltonian corresponds to an effective
%(self-consistent) non-interacting problem. 
The crux of our HMFT is that the approximation is done in the HL where
all possible local OPs are treated on an equal footing and the number
of operators replaced by their mean value is minimized since the
Hamiltonian is quadratic in the symmetry generators. In this way, the
information required is minimal. In mathematical terms, given ${\Cal
O}^\dagger_{\text{\bf ij}} {\Cal O}^{\;}_{\text{\bf ij}} = \langle
{\Cal O}^\dagger_{\text{\bf ij}}\rangle {\Cal O}^{\;}_{\text{\bf
ij}}+{\Cal O}^\dagger_{\text{\bf ij}} \langle {\Cal O}^{\;}_{\text{\bf
ij}}\rangle - \langle {\Cal O}^\dagger_{\text{\bf ij}}\rangle \langle
{\Cal O}^{\;}_{\text{\bf ij}}\rangle+ ({\Cal O}^\dagger_{\text{\bf ij}}
- \langle {\Cal O}^{\dagger}_{\text{\bf ij}}\rangle) ({\Cal
O}^{\;}_{\text{\bf ij}} - \langle {\Cal O}^{\;}_{\text{\bf
ij}}\rangle)$, for an arbitrary bond-operator ${\Cal O}^{\;}_{\text{\bf
ij}}$, the approximation amounts to neglect the latter fluctuations,
i.e.,  ${\Cal O}^\dagger_{\text{\bf ij}} {\Cal O}^{\;}_{\text{\bf ij}}
\approx \langle {\Cal O}^\dagger_{\text{\bf ij}}\rangle {\Cal
O}^{\;}_{\text{\bf ij}}+{\Cal O}^\dagger_{\text{\bf ij}} \langle {\Cal
O}^{\;}_{\text{\bf ij}}\rangle - \langle {\Cal O}^\dagger_{\text{\bf
ij}}\rangle \langle {\Cal O}^{\;}_{\text{\bf ij}}\rangle$. An
important result is that all local OPs are equally treated and,
moreover, symmetries of the original Hamiltonian related to the OPs are
not broken explicitly in certain limits. 

%In a sense,
%this is the {\it best} MF approximation that can be performed, i.e.,
%the best {\it non-interacting} Hamiltonian that approximates the energy
%and correlation functions of the original problem. 

The resulting MF Hamiltonian together with the SW-boson constraint
(with Lagrange multiplier $\lambda$) $\tilde{H}=H_{MF}+\lambda
\sum_{\text{\bf j}} (n_{\text{\bf j}\downarrow} + n_{\text{\bf j}0} +
n_{\text{\bf j}\uparrow})$ reads [10]
$$
\eqalignno{
\tilde{H}&=\! - 2t\bar{n}\! \! \sum_{\langle \text{\bf i},\text{\bf j}
\rangle, \sigma} \! [A (A^\dagger_{\sigma\text{\bf
ij}}+A^{\;}_{\sigma\text{\bf ij}}) - iB (B^\dagger_{\sigma\text{\bf
ij}}-B^{\;}_{\sigma\text{\bf ij}}) ]  - U \sum_{\text{\bf
j}}n_{\text{\bf j}0} + \lambda \sum_{\text{\bf j},\alpha} n_{\text{\bf
j}\alpha}  \cr 
&= \!\!\!\sum_{{\bk}\in \text{\rm BZ}}\![ 
\Lambda_A \ b^\dagger_{{\bk}\text{\sl s}} b^{\;}_{{\bk}0} + \Lambda_B \ 
b^\dagger_{{\bk}\text{\sl s}} b^\dagger_{-{\bk}0}
+\text{\rm H.c.}] + \!\!\!\sum_{{\bk}\in \text{\rm BZ}}[(\lambda-U) 
n_{{\bk}0}+ \!\lambda  (n_{{\bk}\text{\sl s}}+n_{{\bk}\text{\sl a}})],
&(14)} 
%\label{hamiltonMF}
$$
where the sum of momenta $\text{\bf k}$ is performed over the first
Brillouin zone (BZ), $b^\dagger_{{\bk} \text{\sl s}(\text{\sl
a})}=\frac{1}{\sqrt{2}} [b^\dagger_{{\bk}\uparrow}\pm
b^\dagger_{{\bk}\downarrow}]$,  $n_{\text{\bf
k}\alpha}=b^\dagger_{\text{\bf k}\alpha} b^{\;}_{\text{\bf k}\alpha}$
($b^\dagger_{\text{\bf k}\alpha}$'s represent Fourier transformed
modes). $\Lambda_A=-2\sqrt{2} \ t\bar{n}A \gamma_{\bk}$,
$\Lambda_B=2\sqrt{2} \ t\bar{n}B \tilde{\gamma}_{\bk}$, with
$\gamma_{\bk}=2 \sum_{\mu=1}^d \cos(k_\mu)$ and
$\tilde{\gamma}_{\bk}=2\sum_{\mu=1}^d \sin(k_\mu)$.
The resulting self-consistent MF equations are
$$
\cases{ \displaystyle 
A= \frac{\sqrt{2}}{\text{\sl z}N_s} \sum_{{\bk}\in \text{\rm BZ}} 
\gamma_{\bk} \langle b^\dagger_{{\bk} \text{\sl s}}
b^\dagger_{{\bk}0}\rangle_{MF}  \ ,  \cr \displaystyle
B= \frac{\sqrt{2}}{\text{\sl z}N_s} \sum_{{\bk}\in \text{\rm BZ}}
\tilde{\gamma}_{\bk} \langle  b^\dagger_{{\bk} \text{\sl s}}
b^\dagger_{-{\bk}0} \rangle_{MF} \ ,  \cr \displaystyle 
1=\frac{1}{N_s}\sum_{{\bk}\in \text{\rm BZ}} \sum_{\alpha}
\langle n_{{\bk}\alpha}\rangle_{MF} \ .
}  \eqno(15)
$$
On the other hand, it can be shown that the case with vanishing
$B$-ordering is a stable solution. In the zero-temperature limit all
particles condense into the ${\bk =0}$ mode. This condensation corresponds
to the appearance of a superfluid phase of $g$-particles (or $XY$
$S$=1 ferromagnetism in the spin language) and it is manifested in the
non-zero value of $A$:
$$
A=\frac{1}{4\sqrt{2}}\sqrt{16-\left(\frac{U}{\text{\sl z}t\bar{n}}\right)^2}
\ .\eqno(16)
$$
This expression can be easily obtained from Eq. (15) considering that 
at $T=0$ the only non-zero term of the sum is the ${\bk}=0$ one  (i.e.,
all the particles are condensed). From (16) one can immediately see 
that the critical value of $U/t$ for the superfluid-insulator
transition is:  $U_c/t=4{\text{\sl z} \bar{n}}$. 

\vskip 12 truept
\centerline{\bf 2.2  Magnetism and Superfluidity}
\vskip 12 truept

We study now a simple model which displays coexistence and competition
between antiferromagnetism and Bose-Einstein condensation
(superfluidity). The model represents a gas of interacting spin-1/2
Jordan-Wigner bosons with Hamiltonian ($t>0$)
$$
H=t \! \sum_{\langle \text{\bf i},\text{\bf j} \rangle,\sigma} \left (
{\bar{b}}^\dagger_{\text{\bf i} \sigma} {\bar{b}}^{\;}_{\text{\bf j}
\sigma} + \text{\rm H.c.} \right ) + J \sum_{\langle \text{\bf
i},\text{\bf j} \rangle}  (\text{\bf s}_{\text{\bf i}} \cdot  \text{\bf
s}_{\text{\bf j}} -  \frac{\bar{n}_{\text{\bf i}} {\bar{n}}_{\text{\bf
j}}}{4}) + V \sum_{\langle \text{\bf i},\text{\bf j} \rangle} 
{\bar{n}}_{\text{\bf i}} {\bar{n}}_{\text{\bf j}}  - \bar{\mu}
\sum_{\text{\bf j}} {\bar{n}}_{\text{\bf j}} \ ,\eqno(17)
%\label{hamilt}
$$
where $\text{\bf s}_{\text{\bf j}}=\frac{1}{2} {\bar
{b}}^\dagger_{\text{\bf j} \mu} \vec{\sigma}_{\mu \nu}
{\bar{b}}^{\;}_{\text{\bf j} \nu}$ is a $s=\frac{1}{2}$ operator
($\vec{\sigma}$ denoting Pauli matrices). Notice that $H$ is an
extended $t$-$J$-like model of hard-core bosons instead of constrained
fermions [10].  These hard-core bosons could represent three-state
atoms, like the ones used in trapped Bose-Einstein condensates, moving
in an optical lattice potential. For the sake of simplicity we will
only consider the AF, $J>0$, case.

In the HL, $H$ represents a Heisenberg-like Hamiltonian [9] in the
presence of an external magnetic field $\mu'$ ($J_{\mu \nu}=J_{\nu
\mu}$)
$$
H = \sum_{\langle \text{\bf i},\text{\bf j} \rangle} J_{\mu \nu}
{\Cal S}^{\mu \nu}(\text{\bf i}) {\Cal S}^{\nu \mu}(\text{\bf j})
- \mu' \sum_{\text{\bf j}}{\Cal S}^{00}(\text{\bf j}) \ ,\eqno(18)
$$
with $J_{00}=V-J/2$, $J_{01}=J_{02}=t$,  $J_{11}=J_{12}=J_{22}=J/2$,
and $\mu'=\frac{\text{\sl z}}{3}(2V-J/2) - \bar{\mu}$. This 
HL furnishes the natural framework to analyze the symmetries of the
Hamiltonian $H$. There is always an $SU(2)$ spin symmetry generated by
${\Cal S}^{11}-{\Cal S}^{22}$, ${\Cal S}^{12}$, and ${\Cal S}^{21}$.
When $\mu'=0$ and $V=2t$, there are five additional generators of
symmetries related to the charge degrees of freedom. Moreover, if
$J=V=2t$ there is full $SU(3)$ symmetry. For $\mu'\ne 0$, the only
charge symmetry that remains is a $U(1)$ symmetry generated by ${\Cal
S}^{00}$ (conservation of the total charge). In this way the HL,
leading to a unique OP from which all possible embedded orderings are
derived, provides a unified description of the possible thermodynamic
states of the system. Yet, it remains to establish the orderings that
survive as a result of tuning the parameters of the Hamiltonian or
external variables such as temperature and particle filling. 

For arbitrary values of the parameters $J/t, V/t$, we do not know a
priori how to determine exactly the phase diagram of $H$ [12].
The resulting Hamiltonian ($V=2t$ with
no loss of generality) is [10]
$$
H=\! - \! \! \sum_{\langle \text{\bf i},\text{\bf j} \rangle} (
\frac{J}{2} A^\dagger_{\text{\bf ij}}A^{\;}_{\text{\bf ij}} + t\!
\sum_{\sigma=\uparrow,\downarrow}\!
B^\dagger_{\sigma\text{\bf ij}}B^{\;}_{\sigma\text{\bf ij}} ) - \mu 
\sum_{\text{\bf j}}n_{\text{\bf j}0} \ ,\eqno(19)
%\label{hamilton}
$$
where $\mu=\text{\sl z}t -\bar{\mu}$ and the ordering operators
$$
\cases{ 
\;\; A^\dagger_{\text{\bf ij}}= {b}^\dagger_{\text{\bf i}
\uparrow} { b}^\dagger_{\text{\bf j} \downarrow} -
{b}^\dagger_{\text{\bf i} \downarrow} { b}^\dagger_{\text{\bf j}
\uparrow} \cr B^\dagger_{\sigma\text{\bf ij}}= {b}^\dagger_{\text{\bf
i} \sigma} { b}^\dagger_{\text{\bf j}0} - {b}^\dagger_{\text{\bf i}0} {
b}^\dagger_{\text{\bf j} \sigma} \cr} \ ,\eqno(20)
%\label{orders}
$$
which transform as singlets with respect to the generators of $SU(2)$
spin and charge symmetries, respectively: 
$[A^\dagger_{\text{\bf ij}}, {\Cal S}^{12(21)}(\text{\bf i})+{\Cal
S}^{12(21)}(\text{\bf j})]=0= [B^\dagger_{\uparrow(\downarrow)\text{\bf
ij}}, {\Cal S}^{10(20)}(\text{\bf i})$$+ {\Cal S}^{10(20)}(\text{\bf
j})]$.

The resulting MF Hamiltonian reads [10]
$$
\eqalignno{
\tilde{H}&=\! - \! \! \sum_{\langle \text{\bf i},\text{\bf j} \rangle} 
[\frac{J A}{2} (A^\dagger_{\text{\bf ij}}+A^{\;}_{\text{\bf ij}}) +
tB\! \sum_{\sigma=\uparrow,\downarrow}\! (B^\dagger_{\sigma\text{\bf
ij}}+B^{\;}_{\sigma\text{\bf ij}}) ]  - \mu \sum_{\text{\bf
j}}n_{\text{\bf j}0} + \lambda \sum_{\text{\bf j},\alpha} n_{\text{\bf
j}\alpha} \cr 
&=\! \!\! \! \! \sum_{\text{\bf k}\in \text{\rm RBZ}}\! \!\![ \Lambda_A \
b^\dagger_{\text{\bf k} \uparrow} b^\dagger_{-\text{\bf k}+\text{\bf Q}
\downarrow} + \Lambda_B \!\!\!\sum_{\sigma=\uparrow,\downarrow}\!\!
b^\dagger_{\text{\bf k} \sigma} b^\dagger_{-\text{\bf k}+\text{\bf
Q}0}+ \text{\rm H.c.} + \!(\lambda-\mu) n_{\text{\bf k}0}+ \!\lambda 
\!\!\!\sum_{\sigma=\uparrow,\downarrow}\!\! n_{\text{\bf k}\sigma}],
&(21)}
%\label{hamiltonMF}
$$
where the sum of momenta $\text{\bf k}$ is performed over the reduced 
Brillouin zone (RBZ) with AF ordering wave vector {\bf Q}, with 
$\Lambda_A=-2JA \gamma_{\text{\bf k}}$, $\Lambda_B=-4tB
\gamma_{\text{\bf k}}$, with  $\gamma_{\text{\bf k}}=\frac{1}{\text{\sl
z}} \sum_{\vec{\delta}} e^{i \text{\bf k}\cdot\vec{\delta}}$
($\vec{\delta}$ are nearest-neighbor vectors). Note that when $B=0$ in
$H_{MF}$, the $SU(2)$ spin and $U(1)$, ${\Cal S}^{00}$, symmetries are
conserved; the opposite case $A=0$ preserves ${\Cal S}^{10(01)}+{\Cal
S}^{20(02)}$ and  ${\Cal S}^{11}+{\Cal S}^{22}-{\Cal S}^{00}$
symmetries. In  Eq.~(21) we have only considered homogeneous solutions
%Eq.~(\ref{hamiltonMF}) we have only considered homogeneous solutions
[14].
\topinsert
\input psfig.sty
\centerline{\hskip0mm
\psfig{figure=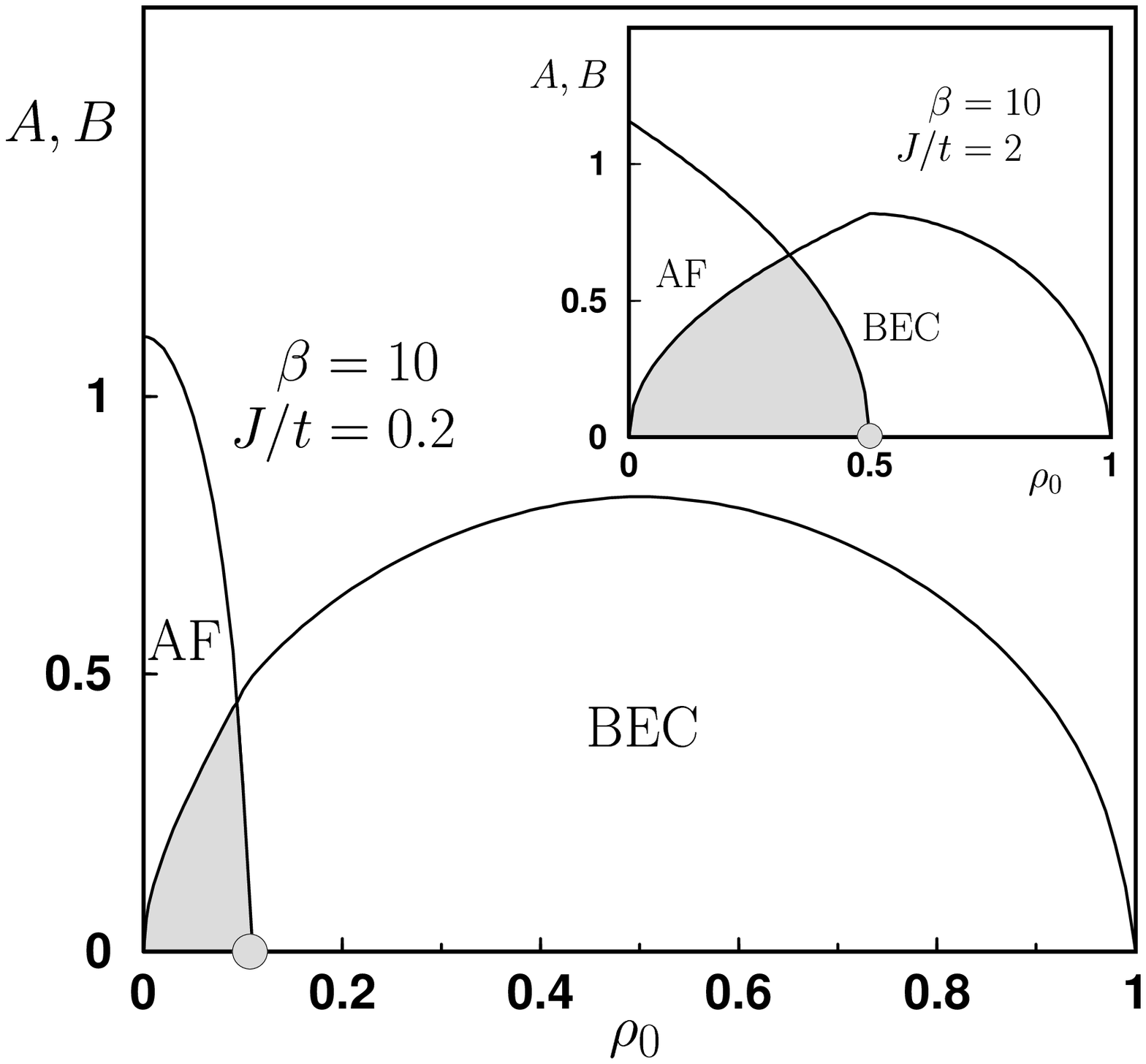,height=8truecm,width=9truecm,angle=0}}
\vskip -0.0truecm 
\noindent
{\bf Figure 1.} 
Order fields $A$ and $B$ as a function of the density $\rho_0$ for 
different values of $J/t$ and inverse temperature $\beta=10$ (in units
of $t^{-1}$). The filled circle on the density axis indicates a quantum
critical point.
\vskip 6truept
\endinsert

The corresponding self-consistent MF equations to solve are
$$
\cases{ \displaystyle 
A= \frac{8}{\text{\sl z}N_s} \sum_{\text{\bf k}\in \text{\rm RBZ}}
\gamma_{\text{\bf k}}  \langle b^\dagger_{\text{\bf k} \uparrow}
b^\dagger_{-\text{\bf k}+\text{\bf Q}\downarrow}\rangle_{MF} \ ,
\cr \displaystyle
B= \frac{8}{\text{\sl z}N_s} \sum_{\text{\bf k}\in \text{\rm RBZ}}
\gamma_{\text{\bf k}} \langle  b^\dagger_{\text{\bf k} \sigma}
b^\dagger_{-\text{\bf k}+\text{\bf Q}0}\rangle_{MF} \ , \cr
\displaystyle 
1=\frac{1}{N_s}\sum_{\text{\bf k}\in \text{\rm RBZ}} \sum_{\alpha}
\langle n_{\text{\bf k}\alpha}\rangle_{MF}\ .
} \eqno(22)
%\label{scf}
$$
We are thus left with a non-interacting system of SW bosons. Now we
follow Colpa [15] and diagonalize para-unitarily the
Hamiltonian matrix $\tilde{H}$. The application of a
homogeneous linear transformation leads to [10]
$$
\tilde{H} = \sum_{\text{\bf k}\in \text{\rm RBZ}} \sum_{i=0}^5
\omega_{i\text{\bf k}} \  \alpha^\dagger_{i\text{\bf k}}
\alpha^{\;}_{i\text{\bf k}} \ ,\eqno(23)
$$
where the mode energies $\omega_{i\text{\bf k}}$ are, at least,
two-fold degenerate. In Fig. 1, we display the orders $A$
and $B$ as a function of $\rho_0=\frac{1}{N_s}\sum_{\text{\bf j}}
\langle n_{\text{\bf j}0}\rangle$ at very low temperature
%\langle n_{\text{\bf j}0}\rangle$ at very low temperature ($\beta=10$)
%and different ratios of the competing interactions $J/t$ for a
and different $J/t$ ratios for a
two-dimensional lattice [16]. The relation between the OPs of the
%original problem, Eq.~(\ref{hamilt}), and $A$ and $B$ is given by
original problem, Eq.~(17), and $A$ and $B$ is given by
$\frac{1}{N_s^2} \sum_{\text{\bf i},\text{\bf j}} e^{i \text{\bf Q}
\cdot (\text{\bf r}_{\text{\bf i}}-\text{\bf r}_{\text{\bf j}} )}
\langle {\bar{b}}^\dagger_{\text{\bf i} \sigma}
{\bar{b}}^{\;}_{\text{\bf j} \sigma'}\rangle_{MF} \propto B^2$,
$\frac{1}{N_s^2} \sum_{\text{\bf i},\text{\bf j}} e^{i \text{\bf Q}
\cdot (\text{\bf r}_{\text{\bf i}}-\text{\bf r}_{\text{\bf j}})}
\langle s^+_{\text{\bf i}}s^-_{\text{\bf j}}\rangle_{MF} \propto A^2$, 
%\begin{eqnarray}
%\begin{cases} \displaystyle 
%\frac{1}{N_s^2} \sum_{{\bf i},{\bf j}} e^{i {\bf Q} \cdot ({\bf
%r_i}-{\bf r_j})} \langle {\bar{b}}^\dagger_{{\bf i} \sigma}
%{\bar{b}}^{\;}_{{\bf j} \sigma'}\rangle_{MF} \propto B^2 \ , \nonumber \\
%\displaystyle
%\frac{1}{N_s^2} \sum_{{\bf i},{\bf j}} e^{i {\bf Q} \cdot ({\bf
%r_i}-{\bf r_j})} \langle s^+_{\bf i}s^-_{\bf j}\rangle_{MF} \propto A^2
%\ ,
%\end{cases} 
%\label{OPss}
%\end{eqnarray}
and justifies the labeling of the phases $\text{\rm AF}$
(antiferromagnet) and $\text{\rm BEC}$ (Bose-Einstein condensate) in
Fig. 1. 

\topinsert
\input psfig.sty
\centerline{\hskip0mm
\psfig{figure=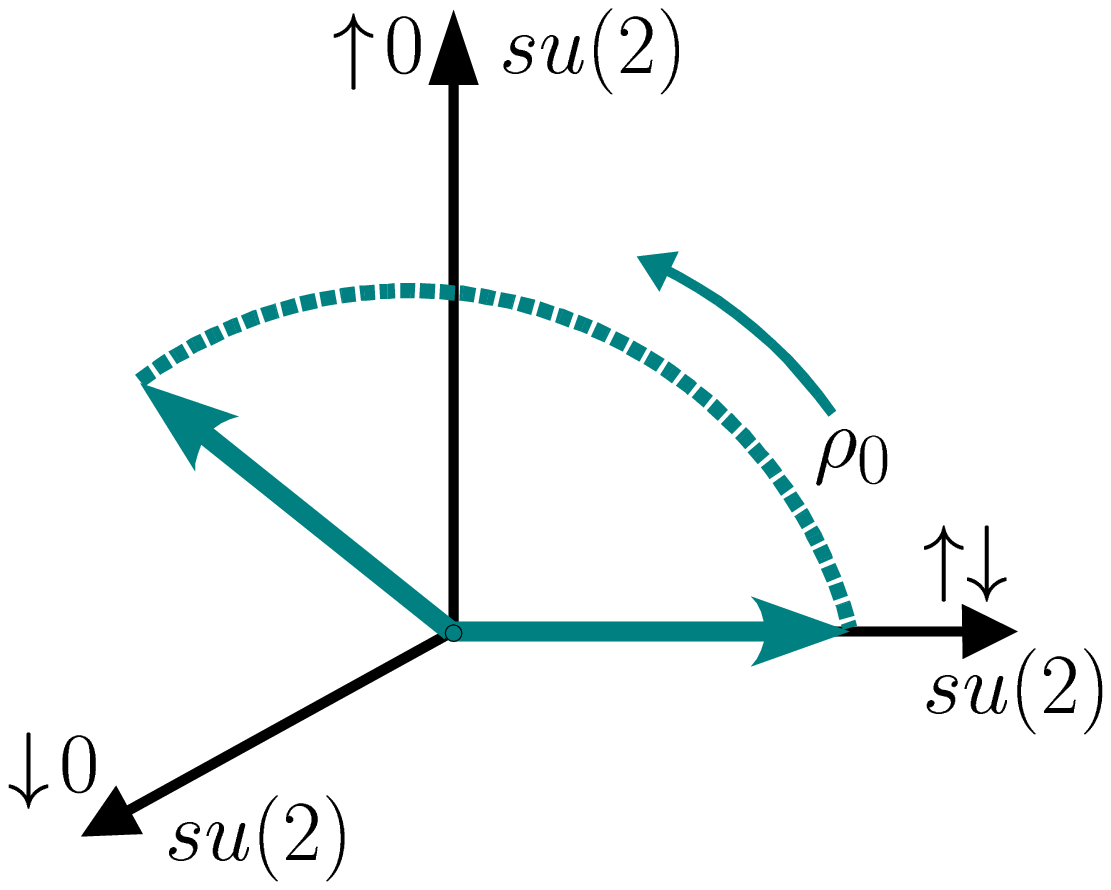,height=6truecm,width=7truecm,angle=0}}
\vskip -0.0truecm 
\noindent
{\bf Figure 2.} 
Schematics of the order parameter change (as a function of
$\rho_0$) with the three $su(2)$ axis describing 3 different $su(2)$
subalgebras of $su(3)$. Note that the plane $\uparrow 0 \ - \!
\downarrow 0$ represents the 5 charge symmetry generators while the
$\uparrow \downarrow$ axis is associated to the remaining three 
generators of magnetism.
\vskip 0truept
\endinsert

A way to qualitatively understand this quantum phase diagram is to look
into the OP space as displayed in Fig. 2. As $\rho_0$ (or
chemical potential) varies from $0$ to $1$, the OP (depicted as
an arrow) moves in OP space. When $\rho_0=0$ the order is purely AF and
the arrow lies on the $\uparrow \downarrow$ axis. For $\rho_0\ne 0$,
the arrow has projections onto the 3 $su(2)$ axis, i.e., the AF state {\it
coexists} with a BEC state. There is a particular critical value of
$\rho_0=\rho_{0c}<1$ for which the AF ordering vanishes and the OP is
purely BEC with the arrow lying in the $\uparrow 0 \ - \! \downarrow 0$
plane. This BEC ordering persists until $\rho_0=1$, where it vanishes.

There are some open issues. One regards the application of the HMFT
approach to study fermionic problems, for example, a Hamiltonian like
%Eq.~(\ref{hamilt}) but where the operators $\bar{b}^\dagger_{\text{\bf i}
Eq.~(17) but where the operators $\bar{b}^\dagger_{\text{\bf i}
\sigma}$ for different modes on a lattice $\text{\bf i}$ anticommute. The
method could certainly be used, however, fermions do introduce a
non-local gauge potential [9] leading to an effective
dynamical frustration which is difficult to handle in a controlled
manner. Another issue concerns the application of the HMFT method when
longer-range interactions are involved. There is already evidence from
%previous work \cite{trumper} on the $J_1$-$J_2$ $SU(2)$  Heisenberg
work on the $J_1$-$J_2$ $SU(2)$ Heisenberg model [17] that
our HMFT will work in those cases as far as homogeneous phases are
concerned. Actually, the SW MF theory introduced by Arovas and Auerbach
[18] in the fundamental representation is a
particular case of our general HMFT. Finally, our methodology exhausts
all broken symmetry instances but it is still quite possible to have
purely topological quantum orders and their corresponding phase
transitions which cannot be described by broken symmetries and 
associated OPs [19] and, thus, are not included in our framework.

\vskip 0.6 truecm

\centerline{\bf 3.  SUMMARY}
\vskip 12 truept

Summarizing, we developed a theoretical framework and a calculational 
scheme to study coexistence and competition of thermodynamic phases in
strongly correlated matter. In our method (given a Hamiltonian modeling
the physical system) the order parameters are not guessed but
rigorously determined from group theoretical considerations as symmetry
generators of a hierarchical language. In this way, the Hamiltonian
operator (which does not necessarily have the full symmetry of the
hierarchical group) is expressed in terms of symmetry generators. Then,
in a non-phenomenological approach dubbed {\it hierarchical mean-field
theory}, we approximated the dynamics (and thermodynamics) treating all
possible local order parameters on an equal footing, i.e., without
preferred symmetry axis. One could say that this procedure follows the
guiding principles of {\it maximum symmetry} and {\it minimum
information}. This allowed us to obtain in a simple manner the phase
diagram of a model problem exhibiting a superfluid-insulator quantum
phase transition and another displaying coexistence and competition
between antiferromagnetism and superfluidity. Combined with an analysis
of the fluctuations (to analyze the stability of the mean-field) one
now has a simple machinery to design phase diagrams.

As can be inferred from our presentation, there are two complementary
aspects to studying competition and coexistence between phase orderings
in strongly coupled quantum systems. One is the direct discovery of the
{\it hidden unity} and subsequent determination of the possible phases
and their transitions, given a Hamiltonian operator modeling the
complex material of interest. This is the aspect we have described in
the present paper. The second aspect, to be discussed in a separate
publication, involves the design or engineering of new states of matter
(i.e., new quantum orders) using the inverse path of logic.
Essentially, the idea consists of tailoring effective Hamiltonians
based upon a general symmetry analysis of the possible orderings one
would like to realize at zero temperature. Tuning the parameters of
these symmetry-based  effective Hamiltonians allows one to move in
parameter space along the previously established orderings. Indeed,
this strategy finds its experimental realization in recent work done on
atomic BEC systems in optical lattices [20], and our approach provides a
unique theoretical guidance to achieve that goal given the possibility
of control and tunability of the interactions of the elementary
constituents (i.e, quantum control). 

%and engineer new states of matter at will, thus laying down a guided
%theoretical combinatorial search for new ground states.

%In his pioneering 1937 paper Landau \cite{landau} assumed the existence
%of an OP and an associated symmetry group, then he proposed that the
%free energy could be expanded near the transition in terms of the OP.

%Some speak about emergent properties and claim that 
%higher organizing principles yet to be discovered are at the root 
%of the problem. Specifically, The claim is that ...

%This work was sponsored by the US DOE under contract W-7405-ENG-36. 
This work was sponsored by the US DOE. GO thanks the US Army Research
Office for travel support. 
%We thank J.E. Gubernatis for a careful reading of the manuscript.

\centerline{\bf REFERENCES}
\vskip 12 truept

\item{[1]}
P.G. Pagliuso {\it et al.}, Phys. Rev. B {\bf 64}, 100503(R) (2001).

\item{[2]}
H. Barnum, E. Knill, G. Ortiz, and L. Viola, quant-ph/0207149.

\item{[3]}
C.D. Batista, G. Ortiz, and J.E. Gubernatis, Phys. Rev. B {\bf 65},
180402(R) (2002).
 
\item{[4]}
G. Ortiz and C.D. Batista, in {\it Recent Progress in Many-body Theories}
(in press).

\item{[5]}
C.D. Batista and G. Ortiz, cond-mat/0207106.

\item{[6]}
L.D. Landau, Phys. Z. Sow. {\bf 11}, 26, 545 (1937).

\item{[7]}
A. Auerbach, {\it Interacting Electrons and Quantum Magnetism} (Springer
Verlag, New York, 1994). 

\item{[8]}
For a general introduction to the standard Bose-Hubbard model one can
consult S. Sachdev, {\it Quantum Phase Transitions} (Cambridge
University Press, Cambridge, 1999), Chap. 10. 

\item{[9]}
C.D. Batista and G. Ortiz, Phys. Rev. Lett. {\bf 86}, 1082 (2001). 

\item{[10]}
Irrelevant constant terms in the transformed Hamiltonians are always
omitted. 

\item{[11]}
Formally, a language is defined by a set of operators each acting
irreducibly on a local Hilbert space of dimension $D$. The language
does not determine the quantum dynamics of the system (which is
determined by the Hamiltonian) but the dynamics can be expressed in
different languages. See Refs.~[4,5].

\item{[12]}
We know that for $J=V=2t=\frac{2\bar{\mu}}{\text{\sl z}}<0$, we can
find the exact ground and lowest energy states of $H$ [3]. 

\item{[14]}
If we are looking for solutions which could break the lattice 
translational symmetry (inhomogeneous), we should consider OPs ${\Cal
S}^{\mu \nu }(\text{\bf k})=\frac{1}{N_s} \sum_{\text{\bf j}} e^{i
\text{\bf k} \cdot \text{\bf r}_{\text{\bf j}}} \ {\Cal S}^{\mu \nu
}(\text{\bf j})$,  for all possible $\text{\bf k}$-vectors. In general,
non-local OPs can be derived (starting from our local classification)
similarly to what has been done in the literature. For instance, for a
%quantum $S=1$ system there are two possible local OPs:  the local
$S=1$ system there are two possible local OPs:  the local
magnetization and spin-nematic parameters. Thus, in the
case of a quantum spin ($S=1$) glass there will be two types of
glasses: One where the local magnetization is frozen at each site and
another where the local spin-nematic order is frozen at each site. In
other words, there will be two Edwards-Anderson OPs. 

\item{[15]}
J.H.P. Colpa, Physica A {\bf 93}, 327 (1978).

%\item{[16]}
%Find a matrix $T^{-1}$ such that $(T^\dagger)^{-1} \tilde{H} T^{-1}
%= \text{\rm diagonal}$,  and $T$ satisfies the para-unitarity condition
%$T^\dagger\Punitary T =  T \Punitary T^\dagger = \Punitary$ with
%$\Punitary=\begin{pmatrix} \one& \Zero \\ \Zero & -\one
%\end{pmatrix}$. The necessary and sufficient condition for an Hermitian
%matrix to be para-unitarily diagonalized (with all diagonal elements
%positive) is that the matrix be positive definite.

%\item{[17]}
%In this case the eigenvalues can be determined in closed analytic form
%($p=0,1,2$)
%$$
%(\omega_{p{\bf k}})^2 \!\!\!\!&=&\!\!\epsilon\!+\!\sqrt{(\epsilon -
%(\lambda-\mu)^2)^2+ 16 \mu \Lambda_B^2 \frac{\lambda-\mu}{3}} \cos\!
%\left(\! \frac{\phi+2p\pi}{3} \!\right) \nonumber \\
%\epsilon &=& \frac{2}{3} (\lambda^2 + \frac{1}{2} (\lambda-\mu)^2 -
%2\Lambda_B^2-\Lambda_A^2)\nonumber \\
%\cos \phi&=&\frac{2\eta(\eta^2+\Lambda_B^2
%(\Lambda_A^2-\Lambda_B^2-\mu^2))-2
%\Lambda_B^2(\Lambda_A^2+\mu^2)}{\sqrt{((\Lambda_B^2-\eta)^2
%-\frac{4}{3}\Lambda_B^2\mu(\mu-\lambda) )^3}} \nonumber\\
%3\eta&=&\Lambda_A^2-\Lambda_B^2+\mu(\mu-2\lambda)\nonumber
%$$

\item{[16]}
$\text{\sl z}=4$, $\text{\bf k}=(k_x, k_y)$,
$\text{\bf Q}=(\pi, \pi)$ and  $\gamma_{\text{\bf k}}=(\cos k_x + \cos
k_y)/2$. 

\item{[17]}
A.E. Trumper, L.O. Manuel, C.J. Gazza, and H.A. Ceccatto, Phys. Rev.
Lett. {\bf 78}, 2216 (1997).

\item{[18]}
D.P. Arovas and A. Auerbach, Phys. Rev. B {\bf 38}, 316 (1988). 

\item{[19]}
X.-G. Wen, Phys. Rev. B (in press). 

\item{[20]}
M. Greiner {\it et al.}, Nature {\bf 415}, 39 (2002).

\end{document}